\documentclass[a4paper,11pt]{article}
\pdfoutput=1 

\usepackage{jcappub} 

\usepackage{amssymb, amsmath, epsfig, natbib}

\newcommand{\bfr}{{\mathbf{r}}}

\newcommand{\bfk}{{\mathbf{k}}}

\newcommand{\calH}{{\cal H}}
\newcommand{\Mpc}{{\rm Mpc}}

\author[a]{Matthew McQuinn}
\author[b]{Martin White}

\affiliation[a]{Department of Astronomy, University of Washington,
Seattle, WA 98195}
\affiliation[b]{Department of Astronomy, University of California,
Berkeley, CA 94720}

\emailAdd{mcquinn@uw.edu}

\title{Cosmological perturbation theory in $1+1$ dimensions}

\keywords{cosmological perturbation theory -- cosmological parameters from LSS -- power spectrum -- baryon acoustic oscillations -- galaxy clustering}

\abstract{
Many recent studies have highlighted certain failures of the standard
Eulerian-space cosmological perturbation theory (SPT).
Its problems include (1) not capturing large-scale bulk flows
[leading to an ${\cal O}( 1)$ error in the 1-loop SPT prediction for
the baryon acoustic peak in the correlation function],
(2) assuming that the Universe behaves as a pressureless, inviscid fluid,
and (3) treating fluctuations on scales that are non-perturbative as if
they were.
Recent studies have highlighted the successes of perturbation theory in
Lagrangian space or theories that solve equations for the effective dynamics
of smoothed fields.  Both approaches mitigate some or all of the
aforementioned issues with SPT.
We discuss these physical developments by specializing to the simplified
1D case of gravitationally interacting sheets, which allows us to
substantially reduces the analytic overhead and still (as we show) maintain
many of the same behaviors as in $3$D.    In 1D, linear-order Lagrangian perturbation theory (``the Zeldovich approximation'') is exact up to shell crossing, and we prove that $n^{\rm th}$-order Eulerian perturbation theory converges to the Zeldovich approximation as $n\rightarrow \infty$.  In no 1D cosmology that we consider (including a CDM-like case and power-law models) do these theories describe accurately the matter power spectrum on any mildly nonlinear scale.  We find that theories based on effective equations are much more successful at describing the dynamics.  Finally, we discuss many topics that have recently appeared in the perturbation theory literature such as  beat coupling, the shift and smearing of the baryon acoustic oscillation feature, and the advantages of Fourier versus configuration space.  Our simplified 1D case serves as an intuitive review of these perturbation theory results.
}

\begin{document}
\maketitle
\flushbottom

\section{Introduction}

\begin{center}
{\it ``[One] grows stale if he works all the time on insoluble problems, and a trip to the beautiful world of one dimension will refresh his imagination better than a dose of LSD.''}\\
~~~~~~~~~~~~~~~~~~~~~~~~~~~~~~~~~~~~~~~~~~~~~~~~Freeman Dyson
\end{center}

Observations of the large-scale matter distribution show structure on a wide variety of
scales.  The organization of galaxies into a beaded, filamentary
``cosmic web'' appears to be the natural outcome of gravitational
instability in a cold dark matter dominated universe acting on an
almost scale-invariant, Gaussian initial matter density field
\cite{zeldovich70, davis85, springel05}.
The process of structure formation is a competition between gravity
and universal expansion and, as such, is a powerful probe of the cosmology \cite{tegmark04, cole05}.  Owing to the spectrum of matter fluctuations in the Universe, smaller scale matter fluctuations on average become nonlinear, collapse, and virialize at earlier times than larger structures, with nonlinear evolution happening coherently on $\lesssim 10~$Mpc scales at present.  The processes that drive the evolution on these scales can only be studied in detail numerically.  However, when ``smoothing'' the cosmic density and velocity field on scales of $\gtrsim 10~$Mpc, perturbative methods may still provide an accurate description of structure formation.

Linear-order Eulerian cosmological perturbation theory has been tremendously useful for understanding the anisotropies in the Cosmic Microwave Background \cite{hu96, planck} and the statistics of present-day matter fluctuations smoothed over $\gg 10~$Mpc scales \citep{2012MNRAS.425..415S}.    The standard formulation of post-Recombination, Eulerian cosmological perturbation theory solves the continuity and Euler equations, treating the matter as a pressureless fluid, and solutions have been found to all orders in the initial matter overdensity \citep{goroff86, makino92,  jain94}.   Zeldovich \cite{zeldovich70} proposed a different perturbative approach that, instead of solving for densities and velocities, solves for the matter displacement field that is consistent with linear Eulerian perturbation theory, the ``Zeldovich approximation''.  These displacements can then be used to predict the density field, providing a better description than linear Eulerian theory into the mildly nonlinear regime \citep{zeldovich70}.  Zeldovich's Lagrangian theory has been extended so that the  displacement can be calculated beyond linear order \citep{buchert92, buchert94,  catelan95,  ehlers97}.  For a review of standard perturbative methods in cosmology see \cite{bernardeau02}.  In the decades since Eulerian perturbation theory and Lagrangian perturbation theory were devised, the concentration has predominantly been on schemes to accelerate the rather slow convergence of these original theories on mildly nonlinear scales by resuming parts of the expansion \cite{mcdonald06, crocce06, matarrese07, matsubara08}.  However, resummation schemes do not address all of the deficiencies of the original perturbation theories, particularly that their solutions are only valid prior to shell crossing and that they treat nonlinear scales as if they were perturbative.

 Recently there has been renewed effort to address these deficiencies by starting with effective equations that depend only on overdensity fields that have been smoothed so that their RMS is less than unity and, hence, perturbative methods can be rigorously applied \citep{baumann12, carrasco14, mercolli14, carroll14, porto13}.  Indeed, \cite{carrasco14} claimed that these effective theories at 2-loop order remain accurate at the percent level to even $k\approx 0.6\,h\,{\rm Mpc}^{-1}$, or $\sim 4^3$ times more modes than where Eulerian perturbation theory (at the low orders that are calculable) does.    A second recent development has been the construction of higher-order Lagrangian theories \citep{PadWhi09,carlson13, white14}, following the realization that Lagrangian perturbation theories fare better at describing the large-scale advection of matter.  Such advection leads to virtually all of the nonlinear evolution of the baryon acoustic oscillations (BAO) seen in the matter correlation function \citep{ESW07, tassev14a}.  Neither of these developments has come without controversy.
Lagrangian theories still make the same uncontrolled approximations as standard Eulerian perturbation theory.  Current calculations in effective theories rely on arguments that reduce the number of diagrams/coefficients \cite{carrasco12, carroll14, senatore14}.  The remaining coefficients are often treated as free parameters that are fit to the measured power.  It is of debate whether the successes of the effective theories relative to previous theories in matching the nonlinear power spectrum owe to this extra freedom. (The counter-argument, which we ascribe to, is that these extra parameters are required for a consistent theory.  However, there is no guarantee that the low-order expansions can be used reliably close to the non-linear scale, the limit    most applications of perturbative methods are most interested in describing.)  It may even be the case that a fully perturbative approach to describe mildly nonlinear scales will never be successful.  A fully perturbative approach appears to be in conflict with (popular) halo models for large-scale structure -- which have had success at describing the nonlinear evolution in the matter field \cite{seljak00, ma00, cooray02, afshordi06}.  In halo models, the beyond-linear-order evolution is largely described by the density profile of halos, which arises from very nonlinear processes that presumably are not captured by any perturbation theory.  This intuition has inspired the recent development of  hybrid models that use both perturbation theory and assumptions about the matter distribution in and around halos \citep{mohammed14, seljak15}.

In this paper, we break from the standard approach of solving cosmological perturbation theory in the full three dimensions, and instead consider it in one spatial dimension (1D).  One spatial dimension corresponds to modes all oriented in the same direction -- breaking the statistical isotropy generally assumed.  Equivalently, it corresponds to the interaction of infinite sheets of matter where the force is independent of distance from each sheet.  Computing the total gravitational force thus amounts to counting the number of sheets to the left and right.  In our cosmological context, these sheets are moving in a Hubble flow relative to one another (and the dimensions traverse to the sheets are also in the same Hubble flow such that the surface density in each sheet scales as $a^{-2}$). Despite this high degree of symmetry, the same dynamical equations (such as the continuity and Euler equation in the standard Eulerian formulation) apply as in 3D and, by extension, 1D cosmological perturbation theories make all of the same assumptions.\footnote{The only assumption that 1D does not make that underlies almost all 3D perturbation theories is that the nonlinear velocity field is curl-free.}
However, both the numerical simulations and analytic calculations are
significantly simplified in 1D.
The three-dimensional $d^3k$ integrals that occur in all 3D calculations
collapse to one-dimensional $dk$ integrals, allowing the computation of
higher-order solutions more easily.  Simulations are able to have much
higher dynamic range in wavenumber in 1D than in 3D for the same memory
and operation count.
[We show in Appendix~\ref{app:PMcode} that to simulate the matter power
spectrum to $1\%$ within a wavenumber range of $\Delta k=0.01~$Mpc$^{-1}$
requires solving for the dynamics of $\sim10^7$ sheets, whereas to do the
same at a wavenumber of $k\sim 0.1~$Mpc$^{-1}$ -- a mildly nonlinear scale
at $z=0$ in our Universe --  with $\Delta k=0.01~$Mpc$^{-1}$ requires
almost $10^{10}$ particles.]
The reduced cost of simulations allows us to test most of the assumptions of different perturbative
approaches on a wide range of cosmological models.  A final (and very important) advantage is that 1D allows us to calculate
the results of both Eulerian and Lagrangian perturbation theory at infinite
order (both yielding the Zeldovich approximation).
Intriguingly, we show that standard Eulerian and Lagrangian perturbation
theories evaluated at infinite order do not yield a correct prediction for
the matter power spectrum at \emph{any} mildly nonlinear scale in \emph{any} cosmology that we consider because these
theories err at describing the dynamics around collapsed structures.

Of course the Universe has three spatial dimensions, so working in 1D loses some crucial aspects of structure formation.  However, while the gravitational dynamics in 1D differ from those in 3D, 1D still shares many features with 3D.  Most trivially, in the limit that the wavenumbers of two modes have much different magnitudes, their coupling has to be the same (up to a geometric factor) in 1D and 3D.  Less trivially, we show that many results that have recently raised excitement in the perturbation theory literature, such as beat coupling \cite{RimHam06}, the smearing and shifting of the BAO peak owing respectively to the RMS matter displacement \cite{tassev14a, white14} and the coupling to large-scale modes \cite{sherwin12}, and the (un)importance of stochastic terms in the nonlinear evolution \cite{carrasco12, mohammed14, seljak15}, have analogues in the 1D case.  Indeed, we find that discussing all of these recent developments in our simplified 1D case serves as a simple and intuitive review of these perturbation theory results.

All of the calculations in this paper will assume Newtonian gravity, similar to most previous beyond-linear order perturbation theory studies.  This approach is traditionally justified because our cosmology is in the limit where the nonlinear scale is much smaller than the horizon scale -- the scale at which General Relativistic corrections apply.  Indeed, the Newtonian calculation correspond to a gauge choice in General Relativity in this limit \cite{matsubara00,chisari11}.  As it will further simplify our analysis, we shall focus throughout on an
Einstein de Sitter background cosmology ($\Omega_m=1$), commenting on the
transition to other cosmologies where appropriate.
Our calculations in the main body of the text are in comoving coordinates, $x$. We use $u$ to denote the peculiar velocity, and $\nabla$ a derivative with respect to $x$.  Finally, most of our expressions will not explicitly show the temporal argument for densities, power spectra, etc. 

This paper is organized as follows.  In section \ref{sec:1Dpert} we introduce our notation
and conventions, then describe the standard
Eulerian and Lagrangian formulations of cosmological perturbation theory (sections \ref{sec:SPT} and \ref{sec:LPT}), proving their
equivalence, and lastly describe the effective field theory of
large-scale structure (section \ref{sec:EFTLSS}), working out examples for a 1D CDM-like cosmology.  We break from this cosmology in Section \ref{sec:powerlaw}, extending our discussion to 1D cosmologies with power-law initial matter power spectra.  Finally, section \ref{sec:1-loop} discusses the lowest order corrections to the matter correlation function and power spectrum in all of these theories; we show that our 1D setting allows many results of perturbation theory to be more simply derived.  Some of the technical details are relegated to a series of appendices.

\begin{figure}
\begin{center}
\epsfig{file=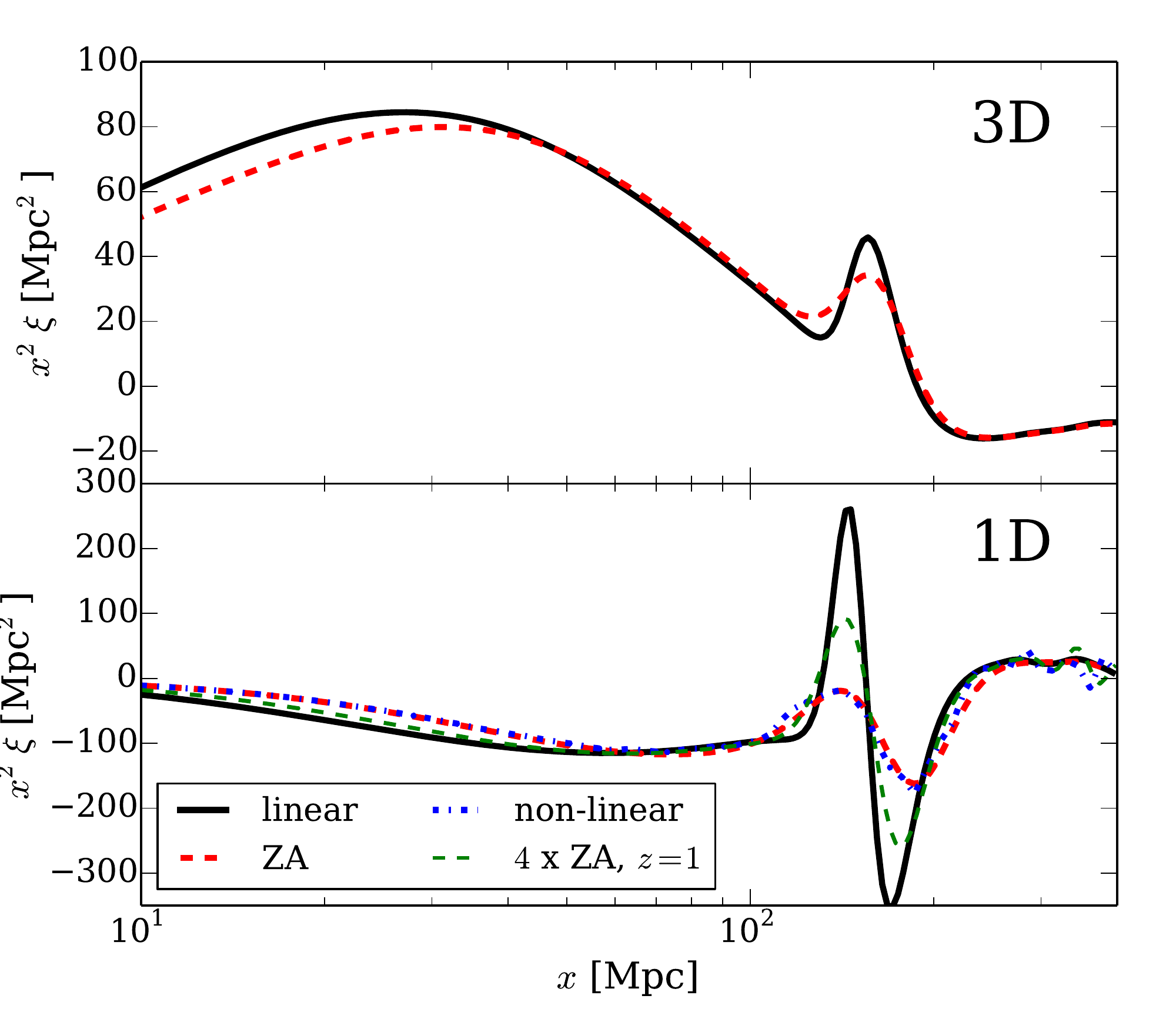, width=10cm}
\end{center}
\caption{Comparison of the correlation function in the concordance $\Lambda$CDM cosmology (top panel) with the correlation function in our 1D CDM-like cosmology (bottom panel).  Both the 1D and 3D cosmologies share the same linear dimensionless matter power spectrum.  All curves are for $z=0$ except the green dashed curve, which shows the Zeldovich approximation at $z=1$ in the 1D cosmology.  In both 1D and 3D, the Zeldovich approximation fares excellently at describing the nonlinear development on $\gtrsim 10~$Mpc scales.  This success can be gleaned in the 1D panel by comparing the Zeldovich Approximation curve with the non-linear curve, which was computed with our $N$-body code.   \label{fig:corrfuncs_comp}}
\end{figure}

\section{Standard perturbation theories in 1D}
\label{sec:1Dpert}

We define the matter overdensity power spectrum, $P(\bfk)$ as
$P(\bfk) \equiv |\widetilde \delta(k)|^2 \times (2\pi)^3/V$,
where $V$ is the volume and $\delta(x) = \rho(x)/{\bar \rho} - 1$, $\rho(x)$ is the matter density
at configuration space position $x$, and tildes indicate Fourier space using
the convention that is standard in cosmology in which the $2\pi$'s only appear
under the $dk$'s. 
We specialize to the case in which modes are directed along a single axis such that  $P(\bfk)$ can be written as
\begin{equation}
  P(k_\parallel, \bfk_\perp; z) = P_{\rm 1d; z} (k_\parallel)
    (2 \pi)^2 \delta^D(\bfk_\perp),
\end{equation}
and we define $\Delta^2 \equiv k_\parallel P(k_\parallel, \bfk_\perp; z)/\pi$
to be the dimensionless power spectrum.
Throughout we shall assume a 1D version of isotropy, ``reflection invariance'',
so that $P_{\rm 1d}(-k)=P_{\rm 1d}(k)$.

We will use a few different parameterizations for the linear-theory power spectrum  in order to explore how different perturbative theories fare.  Mostly we will choose a 1D {linear} perturbation theory power spectrum, $P_L(k; z)$, to have a CDM-like form with
\begin{equation}
  \pi^{-1} k P_{L}(k) = 
    (2\pi^2)^{-1} k^3  P_{\rm CDM}(k; z), 
\end{equation}
where $P_{\rm CDM}$ is the linear theory CDM power spectrum calculated with cosmological transfer function codes such as CAMB,\footnote{\url{http://camb.info}}.  Here, as in what follows, we have dropped the argument for the perpendicular wavevector and the redshift in $P_L$.  This CDM-like form results in the same variance in the density per interval in $k$ as in 3D CDM, as well as the same linear-order parallel RMS displacement.  The top panel in figure~\ref{fig:corrfuncs_comp} shows the linear correlation function in the 3D case, and the bottom panel shows this statistic in the 1D case.  Also shown is the Zeldovich approximation estimate for the nonlinear evolution (which works excellently in both cases; \cite{tassev14a, white14}; section~\ref{sec:zeldovich}).  In 1D, the BAO feature is somewhat sharper, and also the nonlinear evolution at $z=0$ is more substantial.  Empirically, we find that the nonlinear evolution is more similar if we compare the 1D solution at $z=1$ to the 3D one at $z=0$ (compare the thin dashed curve in the bottom panel to the thick dashed in the top panel of figure~\ref{fig:corrfuncs_comp}), an identification used later on.

In section~\ref{sec:powerlaw}, we further consider power-law linear-theory power spectra, parametrized as
\begin{equation}
  (2 \pi)^{-1} k P_{L}(k) = \left( \frac{k}{k_{\rm NL}} \right)^{n+1},
\end{equation}
for $k>0$ and the symmetric form for $k<0$.
The power-law case is particularly
convenient because it admits self-similar solutions in our Einstein de Sitter cosmology, allowing the mapping of
its nonlinear statistical state at one epoch to that at any other epoch. 

\subsection{Standard (Eulerian) perturbation theory}
\label{sec:SPT}

Standard [Eulerian] perturbation theory (SPT) makes the approximation
that the Universe behaves as an ideal fluid, without pressure or viscosity, solving the continuity and pressureless Euler equations
in the presence of gravity
\cite{vishniac83, goroff86, makino92, jain94, bernardeau02}.  In 1D these equations are given by
\begin{eqnarray}
   \partial_\tau {\delta} + \theta &=& -\nabla (\delta u), \label{eqn:cont}\\
   \partial_\tau \theta + {\cal H} \theta + 4 \pi G a^2 \bar{\rho} \delta &=&
    -\nabla (u \nabla u),\label{eqn:euler} 
\end{eqnarray}
where the latter equation is the gradient of the Euler equation with $\theta\equiv \nabla u$, $d\tau = dt/a$, ${\cal H} = a H$, and  $\nabla$ denotes the derivative with respect to the comoving coordinate.  Our 1D symmetry has allowed us to make $u$ a scalar. In Fourier space eqs.~(\ref{eqn:cont}) and (\ref{eqn:euler}) become, 
\begin{eqnarray}
  \partial_\tau {\widetilde \delta}(k) + \widetilde{\theta}(k) &=&
   - \int_{-\infty}^{\infty} \frac{dk'}{2\pi} \frac{k}{k'} \;\widetilde{\theta}(k')
   \; \widetilde\delta(k-k'),\\
  \partial_\tau {\widetilde \theta}(k) + {\cal H}  \;\widetilde{\theta}(k) +
  4 \pi G a^2 \bar{\rho} {\widetilde \delta}(k)  &=&
  - \int_{-\infty}^{\infty} \frac{dk'}{2\pi} \frac{k^2}{2k' (k-k')} \;\widetilde{\theta}(k')
  \; \widetilde\theta(k-k').
\end{eqnarray}
These equations can be solved perturbatively with the ordering
 \begin{equation}
  \widetilde\delta(k, t) = \sum_{m=1}^\infty a^m(t) \widetilde \delta^{(m)}(k),
  \quad
  \widetilde\theta(k, t) =  -{\cal H}(a)  \sum_{m=1}^\infty a^m(t) \widetilde \delta^{(m)}(k),
\end{equation} 
in an Einstein de-Sitter Universe (as considered here), where $a$ is the scale
factor and $\delta^{(m)}$ is the $m^{\rm th}$ order solution for the
overdensity.
(It is common practice to replace $a^m\rightarrow D^m$, where $D$ is the linear
 theory growth factor, in order to treat cosmologies that are not
 Einstein de Sitter.  This replacement is actually exact in 1D.)    In addition to assuming that the universe is a perfect fluid, the SPT expansion assumes that even though $\delta(x) \gtrsim 1$ will often apply, the perturbative solution to these
equations is still valid when smoothing the solution on a scale $R$ such that
$\langle \delta \rangle_R \lesssim 1$.

The solution at each order can be written as
\begin{eqnarray}
  \left(\begin{array}{c} \widetilde \delta^{(n)}(k)\\ \widetilde \theta^{(n)}(k)
  \end{array}  \right) =
  a^{-n} \int \frac{dk_1 ....dk_n}{(2\pi)^{n-1}}
  \delta^D \left(\sum k_i -k  \right)
  \left( \begin{array}{c}F_n(\{ k_i  \})\\ G_n(\{ k_i  \} )\end{array}\right)
  \widetilde \delta_L(k_1)\cdots\widetilde \delta_L(k_n),\label{eqn:deltaF}
\end{eqnarray}
where we have unconventionally kept the expression in terms of the linear overdensity at $a$, $\delta_L$, rather than at $a=1$, and the $F_n$ and $G_n$ obey recurrence relations
\cite{makino92, jain94, bernardeau02}:
\begin{equation}
F_n = (2n+1) X_n + Y_n; ~~~~G_n = 3 X_n + n Y_n. \label{eqn:SPT_rec}
\end{equation}
The argument of each function in these relations is $(k_1, \cdots, k_n)$, and
\begin{eqnarray}
X_n &=& \frac{1}{(2n+3)(n-1)}\sum_{m=1}^{n-1} \frac{k}{K_1} G_m(k_1, ..., k_m) F_{n-m}(k_{m+1}, ..., k_m), \label{eqn:Xn}\\
Y_n &=& \frac{1}{(2n+3)(n-1)} \sum_{m=1}^{n-1} \frac{k^2}{K_1 K_2} G_m(k_1, ... k_m) G_{n-m}(k_{m+1}, ..., k_m), \label{eqn:Yn}
\end{eqnarray}
with $K_1 = k_1 + ... + k_m$, $K_2 = k_{m+1} + ... + k_n$, $k = K_1+K_2$, and $F_1 = G_1 =1$.
For example, the recurrence relations yield $F_2(k, k')=G_2(k,k') = 1 + (k/k' + k'/k)/2$ and $F^{\rm sym}_3(k, k', -k') = k^2/(6k'^2)$, where superscript `sym' indicates the symmetrized form -- $F_n$ averaged over all permutations of its arguments -- which is the form that is easiest to use in calculations.  
Using these solutions for the $\tilde \delta^{(m)}(k)$ to fourth order in $\tilde \delta_L$
(also referred to as `$1$-loop') the matter power spectrum is given by
\begin{eqnarray}
P_{\rm SPT}^{\rm 1-loop}(k) &=& P_{11} + P_{22} +P_{13}, \label{eqn:SPT1loop}\\
P_{22}(k) 
	        & = &  2 \int_{-\infty}^{\infty} \frac{dk'}{2\pi}   F^{\rm sym}_2(k',k-k')^2 P_L(k') P_L(k-k'),\\
                &=&  \int_{-\infty}^{\infty} \frac{dk'}{2\pi} \left\{3 + 4 \frac{k-k'}{k'} + \left(\frac{k-k'}{k'} \right)^2 \right\} P_L(k') P_L(k-k'), \label{eqn:P22}\\
P_{13}(k) & = &  6 P_L(k) \int_{-\infty}^{\infty} \frac{dk'}{2\pi}   F^{\rm sym}_3(k, k', -k') P_L(k'),\label{eqn:P13}\\
&=& -k^2 \eta^2 P_L(k),~~ {\rm where} ~~ \eta^2 \equiv \int_{-\infty}^{\infty} \frac{dk'}{2\pi} \frac{P_L(k')}{k'^2}
\end{eqnarray}
and we have adopted the notation $P_{nm}(k) (2\pi) \delta^D(k - k') = \langle\widetilde \delta^{(n)}(k) \widetilde \delta^{(m)*}(k')\rangle$.\footnote{In deriving eq.~(\ref{eqn:P22}) we have simplified the $P_{22}$ term using the
fact that within the convolution integral we can interchange the arguments of
$F_2$.  Beginning with
\begin{equation}
  F_2^2(k_1,k_2) = \left(1 + \frac{1}{2}
  \left[\frac{k_1}{k_2}+\frac{k_2}{k_1}\right]\right)^2
  =  1 + \left(\frac{k_1}{k_2} + \frac{k_2}{k_1}\right) +
     \frac{1}{4}\left(\frac{k_1}{k_2} + \frac{k_2}{k_1}\right)^2,
\end{equation}
and using the exchange symmetry
\begin{equation}
\frac{1}{4} \left(\frac{k_1}{k_2} + \frac{k_2}{k_1}\right)^2
  \rightarrow \frac{1}{2} + \frac{1}{2} \left(\frac{k_1}{k_2}\right)^2
\end{equation}
in the integral we have $F_2^2\to (3/2)+2(k_1/k_2)+(1/2)(k_1/k_2)^2$. For this way of writing the integral, all of the IR divergences
occur for $k'\simeq 0$, rather than having some also occur at $k\sim k'$,
as occurs on most forms of SPT expressions.
A similar organization is possible in 3D \cite{scoccimarro96,carrasco14b}.}

The 1-loop power spectrum in SPT, $P_{\rm SPT}^{\rm 1-loop}(k)$, receives some contribution from modes
with $k'\gg k$, modes that in many cosmologies are highly nonlinear.  
That SPT shows any sensitivity to nonlinear modes motivates the
effective theories discussed in section~\ref{sec:EFTLSS}.  In the opposite limit in which $k'\ll k$, expanding $P_L(k-k')$ in $k'$, the leading dependence
is\footnote{This is the same as the more often seen $k^2 \eta^2 P_L(k)$ form.
The difference comes because we have written eq.~(\ref{eqn:P22}) to have all
infrared divergences occur for $k' \sim 0$, rather than having some occur at
$k\sim k'$, which leads to $k^2 \eta^2 P_L(k)$ being exactly cancelled.}
\begin{equation}
  P_{\rm SPT}^{\rm 1-loop}(k\gg k') =
P_L(k)  - 2  k \nabla_k P_L(k) \int_{|k'| \ll |k|} \frac{dk'}{2\pi} P_L(k').
\label{eqn:lwc}
\end{equation}
Thus, the sensitivity of $P_{\rm SPT}(k)$ to a long wavelength mode ($k' \ll k$) is via the coupling of its gradient to the large-scale variance, which acts to whiten the power spectrum and smear out features.  The same coupling to large-scale modes holds in the Lagrangian and effective theories discussed later (as all theories discussed here have the same infrared behavior).

  The coupling of $P(k)$ to a long wavelength modes, as given by eq.~(\ref{eqn:lwc}), is different from how the overdensity, $\delta$, couples to such modes.
The coupling of $\delta$ to long-wavelengths modes, and the induced effects
on the power spectrum, is often called `beat coupling'
\cite{RimHam06,HamRimSco06}
or `super-sample covariance' \cite{TakHu13}.  It has been studied
in \cite{meiksin99,RimHam06,TakHu13}.
The coupling of the small-scale mode $\tilde \delta(k)$ with linear overdensity
$\tilde \delta_L(k)$ to the overdensity on some much larger scale $\delta_V$ is
$\tilde \delta(k) =  [1 + 2\delta_{V}] \tilde \delta_L(k)$ in 1D, which differs
only slightly from the 3D result:
$\tilde \delta(\bfk) = [1 + 34\delta_{V}/21] \tilde \delta_L(\bfk)$
\cite{bernardeau02, HamRimSco06, dePutter12, baldauf11}. 
To derive this result note that there are two pieces of the $k'$ integral where the
$2^{\rm nd}$ order contribution to $\tilde\delta(k)$ involves long-wavelength
modes, $\tilde\delta_L(\epsilon)$
(either $|k'|\sim\epsilon$ or $|k-k'|\sim\epsilon$)
and by symmetry both give the same contribution.
Assuming $F_2$ and $\delta_L$ are smooth and can be taken out of the integral
we thus have
\begin{equation}
  \tilde\delta(k)\ni \tilde\delta_L(k) +
  \left[ F_2(+\epsilon,k)+F_2(-\epsilon,k)\right] \tilde\delta_L(k)
  \int_{-\epsilon}^{+\epsilon} \frac{dk'}{2\pi}\ \tilde\delta(k').
\end{equation}
Now noting that the $k'$-integral is just
$\delta_V=L^{-1}\int_{-L/2}^{+L/2}dx\,\delta(x)$ for large $L$ and
$[F_2(-\epsilon,k) + F_2(+\epsilon,k)]\to 2$ as $\epsilon\to 0$,
we obtain the desired result.
This coupling is important for understanding the shift of the BAO peak (section~\ref{sec:1-loop}).

\subsection{Standard Lagrangian perturbation theory}
\label{sec:LPT}

SPT perturbatively expands the RMS relative displacement between points separated by $r$, which at linear order and 1D is
$\propto \int dk\,P(k)/k^2 (1- \cos[k x])$. It so happens that in our 3D Universe and also the 1D CDM-like cosmology the RMS displacement
is the largest perturbative effect at $x\gtrsim 10~$Mpc in the matter correlation function.  Indeed its size, $\sim 15\,$Mpc at $x\sim 100~$Mpc and $z=0$, is comparable to the
width of the BAO peak.  The displacement originates primarily from infrared scales at
which linear theory is a good approximation, suggesting that it should be
fully calculable and should not be treated perturbatively as in SPT.  Lagrangian perturbation theory does not expand in the relative displacement and, hence, has been found to fare much better at capturing the BAO peak in the correlation function
\cite[][and section~\ref{sec:1-loop}]{ESW07,matsubara08,padmanabhan09,PadWhi09,Noh09,tassev13,
carlson13,tassev14a,white14}. 

In its standard form, Lagrangian perturbation theory (LPT) attempts to solve perturbatively the equation \cite{bernardeau02}\footnote{As with SPT, this approach in 3D assumes vorticity is zero and hence only considers the equation for the scalar part of the displacement, a technicality of no importance to our 1D analysis.}:
\begin{equation}
\ddot \Psi(q) + 2 H \dot \Psi(q) = - \nabla \phi(q + \Psi).
\label{eqn:LPT}
\end{equation}
Similar to SPT, LPT solves this equation in powers of the linear density field.  Aside from not expanding in the linear theory displacement, LPT has the same problems as SPT.
 Like SPT, LPT makes the assumption that even though $\delta \gg 1$ on certain scales the solutions are still meaningful when smoothed over a scale $R$ such that $\langle \delta \rangle_R \ll 1$.  LPT is also not valid through shell crossing as infinities appear that invalidate the perturbative expansion as shown below.

The solution to eq.~(\ref{eqn:LPT}) can be used to calculate the real
and Fourier space density fields via 
\begin{eqnarray}
1+\delta_{\rm LPT}(x) &=& \int d q \; \delta^D [x - q - \Psi(q) ] = {\rm det}[1+\nabla_q \Psi]^{-1} \Big |_{x = q + \Psi(q)},\\
                        &=& \int d q \int \frac{dk}{2\pi} \;e^{i k [x - q - \Psi(q)]},\\
\widetilde \delta_{\rm LPT}(k) &=& \int d q \;e^{-i k q}\left(e^{-i k \Psi(q)} -1 \right).\label{eqn:deltakLPT}
\end{eqnarray}
It follows that the power spectrum of the density is
\begin{eqnarray}
P_{\rm LPT}(k) &=&  L^{-1} \int d q_1 dq_2 \; e^{-i k (q_1 - q_2)}  \left(\left \langle e^{-i k [ \Psi(q_1) -  \Psi(q_2) ]} \right \rangle-1 \right), \\
 &=&  \int d q \;  e^{-i k \,q} \left( e^{\sum_{N=1}^\infty (i^N/N!) \langle [k \, \Delta]^N \rangle_c(q)} -1\right),
 \label{eqn:PkLPTcum}
\end{eqnarray}
where $\Delta \equiv \Psi(q_1) - \Psi(q_2)$ and $L$ is the integration ``volume''.  The last line uses the cumulant expansion theorem
$\langle e^{X}\rangle= e^{\sum_{N=1}^\infty\langle X^N\rangle_c/N!}$,
and $\langle [k \Delta]^N \rangle_c(q)$ denotes the $N^{\rm th}$ cumulant
evaluated at $q=q_1-q_2$.

These cumulants can be calculated by expanding the displacement in powers of
$\delta_L$ and solving eq.~(\ref{eqn:LPT}) as in \cite{catelan95}.
However, in 1D all orders beyond linear order are zero!
There are two ways of seeing this.  The more technical approach is to note that in 1D, $1+\delta = {\rm det}[1 + \nabla_q \Psi]^{-1}$ and
$\nabla = (1 +  \nabla_q \Psi)^{-1} \nabla_q$, such that taking the gradient
of  eq.~(\ref{eqn:LPT}), noting that $\nabla^2 \phi = (3/2) H^2 \delta$,
yields
\begin{equation}
\left(1 +  \nabla_q \Psi(q) \right)^{-1} \nabla_q
  \left[ \ddot \Psi(q) + 2 H \dot \Psi(q)\right] =
  \frac{3}{2} H^2 \left(1- {\rm det}[1 +  \nabla_q \Psi(q)]^{-1}  \right).
\end{equation}
Because in 1D ${\rm det}[1 + \nabla_q \Psi] = 1 + \nabla_q \Psi$, the above equation
reduces to 
\begin{equation}
\nabla_q \left[ \ddot \Psi(q) + 2 H \dot \Psi(q)\right]
  = \frac{3}{2} H^2  \nabla_q \Psi(q),
\label{eqn:zel}
\end{equation}
which is a linear equation and, hence, the linear-order solution is
the exact solution of LPT.
This does not mean that higher order cumulants do not contribute to
the true nonlinear power spectrum.
The expression $1+\delta = {\rm det}[1 + \nabla_q \Psi]^{-1}$ becomes infinite
at shell crossing and, hence, LPT only applies up to shell crossing
(which occurs when $\delta \sim 1$). 

A more physical way to see that the lowest order solution is the exact
solution up until shell crossing is from the behavior of the force for
an infinite sheet of matter.  Since the force is directed towards the
sheet but independent of the distance from the sheet, prior to shell
crossing we can replace $\nabla \phi(q + \Psi)$ in eq.~(\ref{eqn:LPT})
with $\nabla \phi(q)$, which results in a linear equation and so the lowest order solution is exact.

\subsubsection{The Zeldovich approximation}
\label{sec:zeldovich}

Linear (or first) order Lagrangian perturbation theory is called
``the Zeldovich approximation'' \cite{zeldovich70,ShaZel89,SahCol95}.
The Zeldovich approximation is the solution to eq.~(\ref{eqn:zel}) and, hence, is exact up to shell crossing.
Since SPT also fails at shell crossing (as it makes a fluid
approximation) the exactness of the Zeldovich approximation suggests
that in 1D LPT should fare at least as well as SPT evaluated to any order.
In fact, we will show that 1D SPT converges to the Zeldovich approximation with increasing order.

The Zeldovich displacement for a particle initially at (Lagrangian)
position $q$ is 
\begin{equation}
  \Psi_{\rm ZA}(q) = \int \frac{dk}{2\pi} e^{i k \,q} \frac{i}{k} \tilde \delta_L(k). \label{eqn:disp}
\end{equation}
Following eq.~(\ref{eqn:deltakLPT}), the overdensity is
\begin{eqnarray}
 \widetilde \delta_{ZA} &=& \int dq\,e^{-ikq} \left( e^{-ik\Psi_{\rm ZA}(q)}-1 \right), \\
  &=& \int dq\,e^{-ikq}\ \sum_{n=1}^{\infty}\frac{[-ik\Psi_{\rm ZA}(q)]^n}{n!}, \label{eqn:ZAexp}\\
  &=& \int \sum_{n=1}^{\infty} \frac{dk_1 \cdots dk_n}{(2\pi)^{n-1}}   \delta^D \left(\sum_{i=1}^n k_i -k  \right) F_n^{\rm sym}(k_1, \cdots, k_n) \tilde \delta_L(k_1) \cdots \tilde \delta_L(k_n),\label{eqn:ZAexpFn}
\end{eqnarray}
where  for $F_n$ it follows from eqs.~(\ref{eqn:disp}), (\ref{eqn:ZAexp}), and (\ref{eqn:ZAexpFn}), and from Appendix~\ref{ap:proof} for $G_n^{\rm sym}$, that \cite{GriWis87}
\begin{equation}
  F_n^{\rm sym}(k_1,\cdots,k_n) = G_n^{\rm sym}(k_1,\cdots,k_n) 
  = \frac{1}{n!}\ \frac{k^n}{\prod_{i=1}^n k_i},
  \label{eqn:FnLPT}
\end{equation}
with $k=k_1+\cdots+k_n$.    We have labelled the Zeldovich kernel as $F_n^{\rm sym}$ and $G_n^{\rm sym}$ to identify them with the symmetric kernel in SPT (eqs.~\ref{eqn:SPT_rec} and \ref{eqn:deltaF}).  This identification is proven in Appendix~\ref{ap:proof}.  Hence, {\it as $n\rightarrow \infty$ the 1D SPT solution converges to that in LPT (the Zeldovich approximation).}  The result that SPT converges to LPT is often taken for granted in perturbation theory studies (in 3D), as \cite{matsubara08} and \cite{rampf12} have shown respectively that to third and fourth order in $\delta_L$ the two theories agree. 
  This result that the expansion of LPT yields the same as SPT is nonetheless surprising:  The dynamical equations for the two theories are quite different (with LPT allowing streams of matter unlike SPT).  Nevertheless, the perturbative solutions to these equations are identical.\\

All statistical quantities in the 1D Zeldovich approximation can be calculated from the variance of the differential displacement between two points separated by distance $q$:
\begin{equation}
  \sigma^2(q) =
  \langle\left[\Psi_{\rm ZA}(0)-\Psi_{\rm ZA}(q)\right]^2\rangle
  =  \int_0^\infty \frac{dk}{\pi} \frac{2\,P_L(k)}{k^2}
  \left(1 - \cos[k\,q] \right).
\label{eqn:sigmaq}
\end{equation}
For example, the Zeldovich approximation power spectrum follows from
eq.~(\ref{eqn:PkLPTcum}) \cite{taylor93, schneider95, fisher96}:
\begin{eqnarray}
  P_{\rm ZA}(k) &=&  \int d q \;  e^{-i k \,q}
  \left( e^{-k^2\sigma^2(q)/2} -1\right), \label{eqn:PkZeld}\\
  &=& \int dq\, e^{i k \,q} \sum_{n=1}^{\infty}
  \frac{\left[-k^{2} \sigma^{2}(q)  \right]^n}{2^n n!},
  \label{eqn:Pksigexp} 
\end{eqnarray} 
where we have used that the displacements are Gaussian random and so
only the second cumulant is nonzero and equal to $\sigma^2$, and the second line simply expands the exponential.  Since LPT has identical $F_n$ as in SPT, it follows that the $n=0$ term in eq.~(\ref{eqn:Pksigexp})  yields linear theory, the $n=2$ term is identical to 1-loop SPT, and so on.  With this identification between order in $\sigma^2$ and the loop in SPT, once $\sigma^2$ is pre-tabulated eq.~(\ref{eqn:Pksigexp}) can be used to calculate the
SPT contribution to any order in 1D with a single integral
over $q$ (and indeed, for numerical evaluation, a sum of Fast Fourier Transforms).  In addition, the correlation function is just the Fourier transform of the power spectrum 
and given by \cite{carlson13}
\begin{equation}
  1 + \xi_{\rm ZA}(r_\parallel) =
  \int \frac{dq}{\sqrt{2\pi}\, \sigma(q)}
  \exp\left[-\frac{(q-r_\parallel)^2}{2\sigma^2(q)} \right].
  \label{eqn:xiZeld}
\end{equation}
Despite its apparent simplicity, because of the $q$ dependence of
$\sigma$ the RHS has to be evaluated numerically with rare exception.  Equation~(\ref{eqn:xiZeld}) can also be expanded in $\sigma^2$ and its derivatives, with each increasing order being one higher loop in SPT (Appendix~\ref{app:LPTtoSPT}).

\subsubsection{Common resummation schemes}

In the decades since standard Eulerian and Lagrangian perturbation theory were devised, the concentration has predominantly been on schemes to accelerate the convergence of these original perturbation theories by resuming terms in the original expansions \cite{mcdonald06, crocce06, matarrese07, matsubara08}.   Since these resummation theories are based on the same dynamical equations as standard perturbation theories, they still rest on the standard expansion converging to a meaningful result -- one question we aim to address by studying 1D dynamics.  Their infinite order solutions are indeed the same as this limit in SPT and LPT.

The ``improvement'' of many of these schemes is in essence to write eq.~(\ref{eqn:Pksigexp}) without expanding the zero lag term
$\eta = \langle \Psi^2(q)\rangle$ so that in 1D
\begin{eqnarray}
  P_{\rm resum}(k) &=& e^{-k^2 \eta} \int dq\, e^{i k \,q} \sum_{n=1}^{\infty} \frac{1}{n!}\left[-k^{2} \langle \Psi(q') \Psi (q' + q)\rangle  \right]^n, 
\end{eqnarray} 
where at ``linear'' order (the $n=1$ term) $P(k) = \exp[-k^2\eta^2] P_L$.
This form appears in the Lagrangian theory of Matsubara \cite{matsubara08}
and an analogous form in Renormalized Perturbation Theory
\cite[RPT,][]{crocce06}.
In RPT, this zero-lag exponential suppression factor was championed as a feature of the theory that it ``shields'' from
small wavelength modes.
However, this exponential suppression is artificial, occurring because only part of
$\sigma^2(q)$ has been kept exponentiated; the other part that tends to
cancel it has been expanded (note $\sigma^2(q)\rightarrow0$ as $q\rightarrow 0$).
Physically this expansion breaks the Galilean invariance of the theory at
every finite order: small-scale perturbations do not depend on the
properties of large-scale flows at any order in SPT or LPT
\cite[this becomes evident in later expressions]{tassev14a,carlson13,porto13};
however, the $\exp[-k^2\eta^2]$ term introduces such a dependence.
While $P(k) = \exp[-k^2\eta^2] P_L(k) $ reproduces the suppression of the
BAO in the correlation function in the concordance cosmology, this success
is largely a coincidence.
For example, for power-law 3D cosmologies with $-3 < n_{3d} < -2$, the
non-exponentiated form leads to the nonsensical result $\eta=\infty$
and, hence, $P(k)=0$ at finite order even though LPT is still convergent.
  
\subsection{Comparison of standard perturbation theories}

\begin{figure}
\begin{center}
\epsfig{file=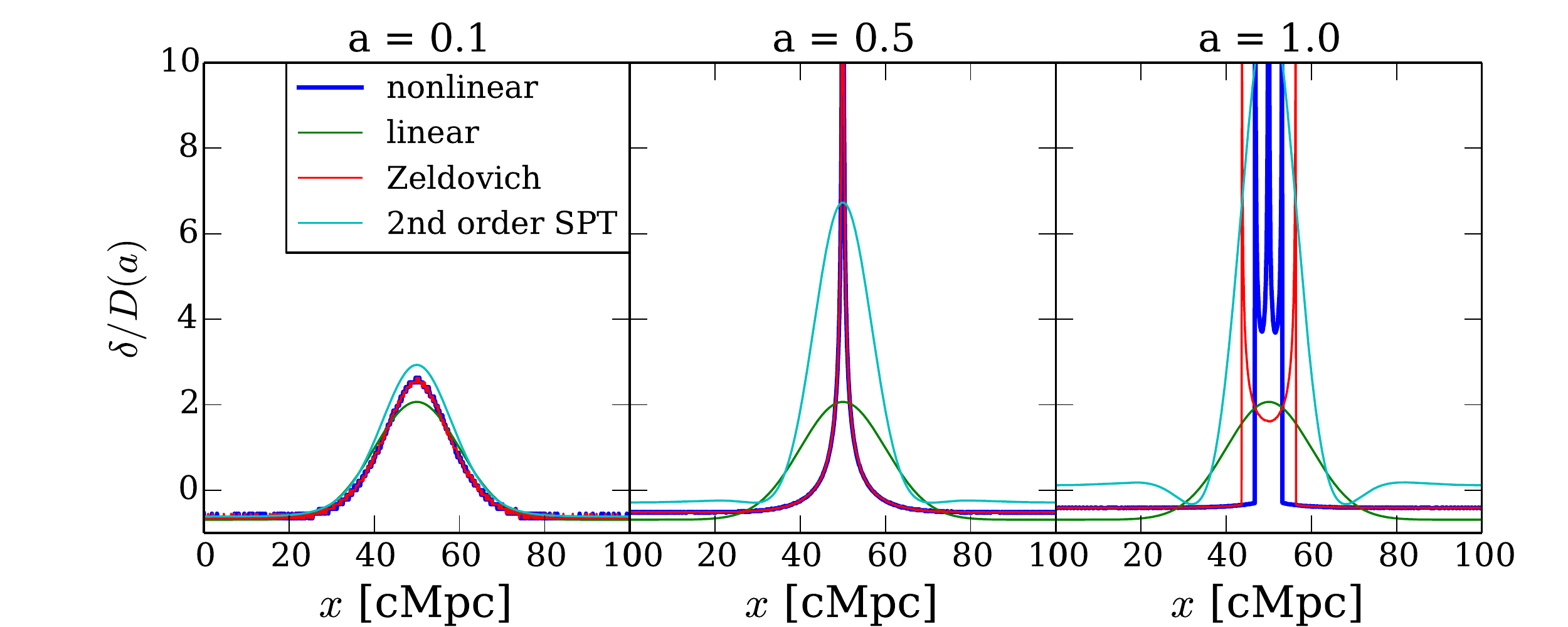, width=15cm}
\epsfig{file=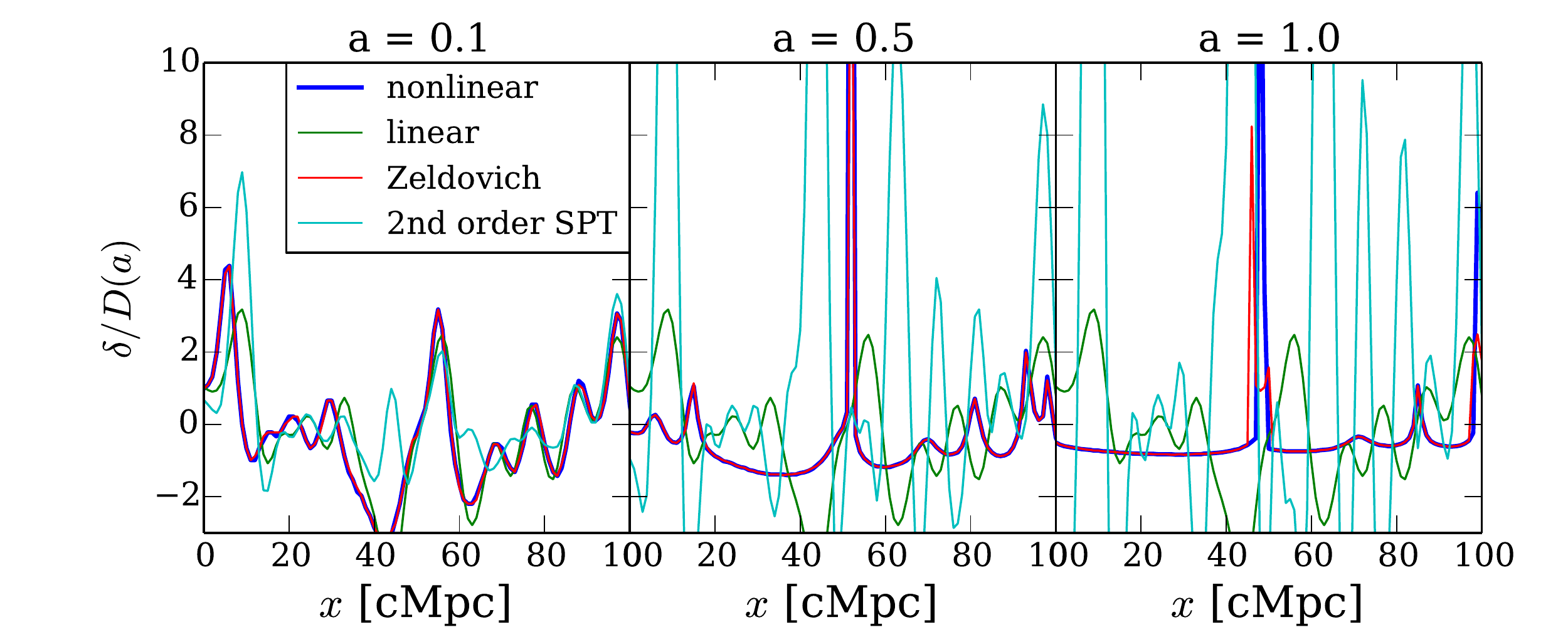, width=15cm}
\end{center}
\caption{{\it Top panel:} The evolution of a Gaussian perturbation that in linear theory has standard deviation of $10~$Mpc and an amplitude at peak of $\delta_L = 2$ for $a=1$.  The ``nonlinear'' curves are computed using our particle mesh code, whereas the other curves are the predictions of Eulerian linear theory, the Zeldovich approximation, and second order Lagrangian perturbation theory.  {\it Bottom panel:}  The same calculations are shown but for a realization of our CDM-like cosmology and where we have additionally damped the linear power spectrum by $\exp[-5 k^2]$ to reduce features.  This frame shows only $1\%$ of the $10^4$~Mpc box used to calculate these fields.  
  \label{fig:gaussian_pert}}
\end{figure}

Here we provide a physical sense for how the aforementioned theories behave before we turn to newer effective theories.  Since we showed that SPT, LPT, and resummation schemes converge to the same result, in a sense these are all the same theory.  Figure~\ref{fig:gaussian_pert} compares the predictions of linear theory, the Zeldovich approximation (equivalent to LPT at any order), 2$^{\rm nd}$ order SPT, and the full nonlinear solution calculated with our $N$-body code (section~\ref{app:PMcode}).  We show the predicted evolution at $a=0.1,~0.5$ and $1$ for both a Gaussian perturbation with initial standard deviation of $10~$Mpc that achieves $\delta_L = 2$ at its peak when $a=1$ (top panel) as well as our CDM-like cosmology where we have additionally damped the linear power spectrum by $\exp[-5 k^2]$ to eliminate small-scale structures (bottom panel).  Especially in the bottom panel of figure~\ref{fig:gaussian_pert}, we see that $2^{\rm nd}$ order SPT performs very poorly relative to the full nonlinear solution.  Even smoothing the SPT field on $\sim 10\,$Mpc scales -- scales at which this theory is frequently used -- will not cause it to fair nearly as well as LPT.  Both panels in figure~\ref{fig:gaussian_pert} show that the Zeldovich approximation performs excellently in most locations.  However, in nonlinear locations where shell crossing has occurred the structures in the Zeldovich approximation are more extended than in the simulations.  The Zeldovich approximation also tracks the advection of structures much better than either linear theory or 2$^{\rm nd}$ order SPT, which is very apparent in a movie of the temporal evolution in the case shown in the bottom panel.  In such a movie, the linear theory structures are fixed in position, whereas the structures in LPT and in the simulation advect substantially across the $100~$Mpc frame.

\begin{figure}
\begin{center}
\epsfig{file=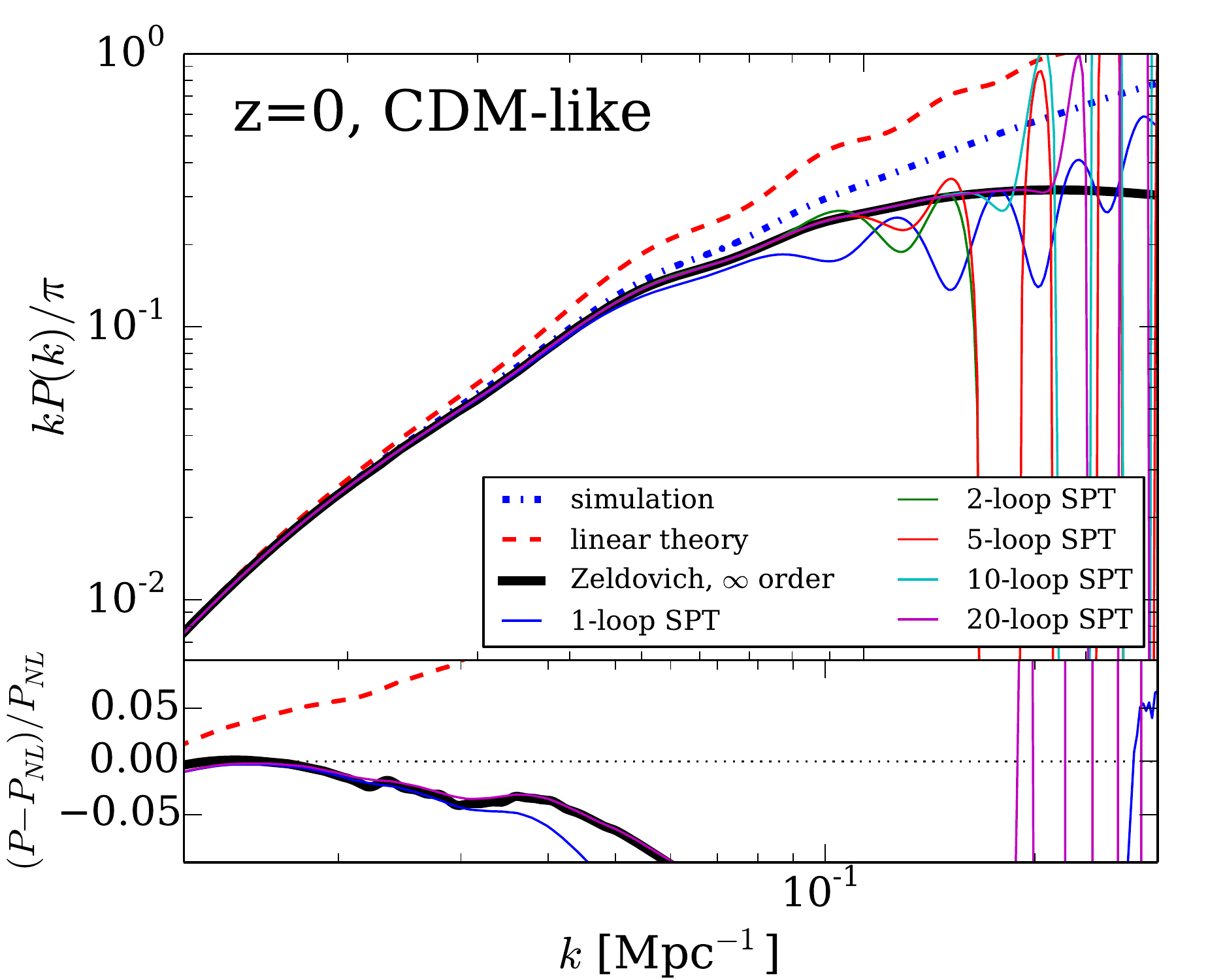, width=12cm}
\end{center}
\caption{{\it Top panel:} $z=0$ matter overdensity power spectrum in our 1D CDM-like model calculated analytically using linear theory, the Zeldovich approximation (LPT at any order), and SPT to the specified order in the overdensity.  Note that infinite order SPT yields the Zeldovich approximation (which is exact up to shell crossing).   The dot-dashed curve is the nonlinear evolution calculated from a simulation using this cosmology.  None of the calculations converge to the simulation result except on scales where linear theory holds.  {\it Bottom panel:} Shown are the fractional residuals with the nonlinear calculation of select perturbation theory curves from the top panel.\label{fig:SPT}}
\end{figure}

A more quantitative comparison of these theories can be done using the power
spectrum. Figure~\ref{fig:SPT} shows this calculation for our CDM-like 1D
model and for various orders in SPT, including the 20-loop power spectrum
(i.e.~the solution to order $P_L^{21}$!):
As the SPT order increases, the solution tracks the Zeldovich approximation prediction to higher and higher wavenumbers.
Each higher order in the expansion has terms that appear with a higher
power of $k$ than the previous order, which manifests in the high frequency
oscillations that are seen at $k>0.1~$Mpc$^{-1}$ in the $\geq 5$-loop order calculations.
One difference between the 1D and 3D cases is that 1D is guaranteed to
converge on all scales to the Zeldovich approximation, whereas in 3D there
is no reason to believe that the series converges to anything
finite on scales where $\delta \gtrsim 1$.
However, even though the series converges in 1D, it does not converge to the
result of the $N$-body simulation given by the dot-dashed curve in figure~\ref{fig:SPT}.
Somewhat surprisingly, even though the Zeldovich approximation describes the
position of structures in figure~\ref{fig:gaussian_pert} very well, the power
spectra differs considerably from the simulation starting at $k\sim 0.02~$Mpc$^{-1}$.
Because the Zeldovich approximation is exact until shell crossing, this bias
must owe to the deficiencies of the perturbation theory once shell crossing
occurs.
This shows that small errors on $\sim 1~$Mpc scales propagate to much smaller $k$, highlighting one of the difficulties of a successful perturbation theory.

\subsection{Effective theories of large-scale structure}
\label{sec:EFTLSS}

We have already noted that the traditional perturbation theory expansions treat nonlinear scales as if
they were perturbative and treat the cosmological matter field as a perfect fluid.  The goal of the effective field theory of
large-scale structure is to overcome these deficiencies by formulating a theory for the dynamics of perturbative (large-scale) modes.  These theories attempt to write dynamical equations for the smoothed matter overdensity and velocity fields, where the smoothing is on large enough scales that perturbation theory can be rigorously applied \cite{baumann12,carrasco12,pajer13,manzotti14,carroll14, porto13}.
The effects of the short-range dynamics on the smoothed fields is encapsulated in a series of parameters.  
  In the next subsection we detail the effective theory approach in an Eulerian context.  (See also Appendix \ref{ap:EFTsymmetry} for a description of an approach that derives the effective equations based just on symmetries, summarizing \cite{manzotti14}, although these symmetries are also taken into account in the standard approach.)  
  A pseudo--Lagrangian effective theory, and a comparison to simulations are treated in ensuing subsections.

\subsubsection{The (Eulerian) effective field theory of large-scale structure}
\label{ss:EFTLSS}

The following summary of the effective field theory of large-scale structure (EFTLSS) involves several steps, but the final result at 1-loop order is simple and follows from adding a new term to the Euler equation that at lowest order in $\delta/u$ is not forbidden by symmetry (and physically corresponds to including effective sound and viscosity terms).  Let us define the smoothing operation on
field $X$ as
\begin{equation}
  X_l  = \int dx \, W_\Lambda(x-x') X(x'),
\end{equation}
where $W_\Lambda$ is a window function with comoving width $\Lambda^{-1}$ chosen
such that $\Lambda<  k_{\rm NL}$.  Here $k_{\rm NL}$ is the ``nonlinear'' wavenumber below
which $\delta < 1$ such that the smoothed field can be treated perturbatively.  
The EFTLSS as presented in \cite{baumann12, carrasco12, pietroni12} assumes that
dynamically the universe is a system of collisionless particles.
The equations for the dynamics of the smoothed fields are  
\begin{eqnarray}
  \partial_\tau {\delta}_l + \theta_l   +\nabla ( \delta_l u_l ) &=&- \nabla [u (\delta - \delta_l)]_l,\label{eqn:effcont}\\
\partial_\tau \theta_l + {\cal H} \, \theta_l  + 4 \pi G a^2 \delta_l +\nabla (u_l \nabla u_l)  &=&  -\bar \rho^{-1} \nabla^2 \tau_\Lambda  -  \nabla \partial_\tau  \{[u (\delta - \delta_l)]_l (1-\delta_l)\}\label{eqn:effnavierstokes}\\
&&   - \nabla^2 \left( u_l  [u (\delta - \delta_l)]_l \right) -  \calH \nabla \{ [u (\delta - \delta_l)]_l (1-\delta_l) \},\nonumber
\end{eqnarray}
both valid to ${\cal O}(\delta_l^2,  \Lambda^{-2})$ for the short-wavelength sensitive terms on the RHS, the order required for a 1-loop calculation as demonstrated later.  Here
 \begin{equation}
 \tau_\Lambda = [\sigma_{u,l}^2 - \rho_l u_l^2] - \frac{1}{8 \pi G a^2}\left[[(\nabla \phi)^2]_l - (\nabla \phi_l)^2\right],
 \label{eqn:tau}
\end{equation}
$\rho_l$, $u_l$, $\sigma_{u,l}$, and $\bar \rho$ are the large-scale density,
peculiar velocity, peculiar momentum dispersion, and mean matter density.  Appendix \ref{app:stresstensor} derives these equations in 1D for a set of
gravitationally interacting sheets, starting from the Vlasov-Poisson
equation.    Both eq.~(\ref{eqn:effcont}) and eq.~(\ref{eqn:effnavierstokes}) are similar to the continuity and Euler equation used in SPT
except (1) the equations are in terms of the smoothed (or long wavelength) fields and (2) there are extra terms that are sensitive to small-scale perturbations.  The extra terms in our equations are different from earlier attempts, such as \cite{baumann12, carrasco12}, in which only the ``stress tensor'' term, $\tau$, appeared on the RHS of the Euler equation and no terms on the RHS of the continuity equation.  This difference only owes to trying to formulate our equations in terms of the Eulerian velocity rather than the less physically meaningful velocity that appears in these other papers (see Appendix \ref{app:stresstensor}).  This reformulation does \emph{not} have any impact on the final results, but we do use this formulation when we estimate the free parameters of this theory from simulations in Appendix~\ref{sec:estcs}.


To solve perturbatively the EFTLSS equations
(\ref{eqn:effcont}--\ref{eqn:effnavierstokes})
requires calculating the retarded Green's function of the linear part of
these equations (e.g., \cite{crocce06, carrasco12}).
First, it is simplest to reformulate these equations as a second-order
differential equation in just $\delta_l$:
\begin{eqnarray}
-a^2 {\cal H}^2(a) \partial_a^2 \delta_l -a \left(2 {\cal H}^2(a) + a {\cal H}(a) d{\cal H}(a)/da \right)   \partial_a \delta_l  + 4 \pi G \bar{\rho} a^2 \delta_l  = \nonumber \\ \left( a {\cal H} \partial_a + {\cal H} \right)  \nabla (\delta_l u_l) -\nabla (u_l \nabla u_l) \nonumber \\
\underbrace{  -\bar \rho^{-1} \nabla^2  \tau_\Lambda + \nabla a \calH \partial_a ( [u(\delta - \delta_l)]_l \delta_l) - \nabla^2 \left( [u (\delta - \delta_l)]_l u_l \right)- \nabla \calH [u (\delta - \delta_l)]_l (1-\delta_l)}_{-\bar \rho^{-1} \nabla^2 X_\Lambda} 
,
\label{eqn:nonlinear}
\end{eqnarray}
where the final line, which we define as $-\bar \rho^{-1} \nabla^2 X_\Lambda$ (such an $X_\Lambda$ exists), are the terms that do not appear in SPT and depend on short-wavelength modes.\footnote{To eliminate $u_l$ so the equation is just in terms of $\delta_l$, we would need to continually plug in
$u_l = \int \theta_l = -\int [\partial_\tau \delta_l + \nabla(\delta_l u_l) + \nabla [u (\delta - \delta_l)]_l]$
into the RHS of the equation.}
The Green's function, $G(a, a')$, of the linear part of this equation satisfies
\begin{equation}
-a^2 {\cal H}^2(a) \partial_a^2 G(a, a') -a \left(2 {\cal H}^2(a) + a {\cal H}(a) d{\cal H}(a)/da \right) \partial_a G(a, a')  + 4 \pi G \bar{\rho} a^2 G(a, a') = \delta^D(a - a').
\end{equation}
The retarded Green's function can be written as a linear
combination of the growing ($\propto a$) and decaying
($\propto a^{-3/2}$) mode solutions with a Heaviside $\theta^H$ function
enforcing the causality-condition $G(a, a')=0$
for $a<a'$:\footnote{One can construct a Green's function from the solutions
of the homogeneous equation by taking the linear combination of the growing and decaying mode that vanishes
at $a=a'$ and then by multiplying this combination by $\theta^H$.}
\begin{equation}
 G(a,a') = \theta^H(a-a')~\frac{2}{5} \calH_0^{-2}
           \left[\left(\frac{a'}{a} \right)^{3/2} -\frac{a}{a'} \right].
\end{equation} 

The aim is to be able to solve eq.~(\ref{eqn:nonlinear}) perturbatively.
To do so, we expand all new small-scale-sensitive terms (which are all not in SPT and hence unique to EFTLSS) in the long-wavelength fields,
$\delta_l$ and $u_l$.\footnote{Up until this expansion, the coarse-grained theory of \cite{pietroni12, manzotti14} is identical to the EFTLSS approach of \cite{carrasco12}.  Instead of expanding, this theory measures $ \tau_\Lambda$ using simulations.}
This expansion should depend on all combinations that do not break translational
invariance and isotropy:
 \begin{equation} 
X_\Lambda = p_{\rm eff}+ \bar \rho c_s^2 \delta_l - \bar \rho \frac{c_v^2}{{\cal H}} \nabla u_l + J(x, t) + ...,
\label{eqn:X}
\end{equation}
where we have also only kept terms that are
lowest order in $\nabla^n$.
Here, we use the same notation as \cite{baumann12} for the coefficients of this expansion --  
$p_{\rm eff}$, $c_s$, and $c_v$ -- because, as shown there, there are analogues between these parameters and pressures, sound speeds and viscosities, and the Navier-Stokes equation for an imperfect fluid, especially when considering only the $\tau_\Lambda$ component of $X_\Lambda$.  Indeed, we will refer to the effective correction to the power spectrum we derive as owing to an effective sound speed. In addition, $J$ is a stochastic component that does not correlate with the long-wavelength fields.  
  These parameters, which encapsulate the short-wavelength physics
that has been ``integrated out'', depend on both $a$ and $\Lambda$.

 There is one complication with our derivation so far (see also Appendix~\ref{app:stresstensor}).  Long-wavelength modes depend on the small-wavelength modes at all previous times (e.g., at all times along a trajectory; \cite{carroll14} and maybe even on $\partial_t^m$ of these modes).  However, the expansion of $X_\Lambda$ in terms of long wavelength modes (eq.~\ref{eqn:X}) assumed locality in time:  that the rate of change of ($\delta_l$, $u_l$) at time $t$ only depends on short-wavelength modes at $t$.  Fortunately, at 1-loop order the final result is equivalent if we treat eq.~(\ref{eqn:X}) as the fundamental expansion as non-locality in time can be absorbed into the time dependence of $c_s$, $c_v$, and $J$.  These complications make the predictions of the previous $>1$~loop EFTLSS calculation depend on an unknown temporal response kernel \cite{carrasco14},\footnote{Different terms that contribute at 2-loop order have different integrals over this response and so, unlike the 1-loop case (where it can be absorbed into a  single coefficient), the form of the temporal response affects the $k$-dependence of the power spectrum.} although it is quite possible that future formulations will resolve this deficiency \cite{mirbabayi14}.

We follow \cite{carrasco12} and treat all terms in $X_\Lambda$, including
$c_s^2 \delta_l$, as nonlinear terms.  The motivation for this will be discussed shortly, but boils down to, e.g., $c_s^2 \delta_l$ serving to cancel
the UV sensitivity of terms that scale as $\delta_L^3$ and so should be thought of as a third order term.  To generate the nonlinear solution to the dynamical equations requires integrating $G(a, a') $ times the nonlinear terms that appear in eq.~(\ref{eqn:nonlinear}).  Since the nonlinear terms depend on unknown fields $\delta_l$ and $\theta_l$, a perturbative solution is required.  We adopt the standard perturbation theory approach, first solving the smoothed linear theory differential equation, yielding $\delta_{L, l}$.  We then feed this solution into the quadratic terms to solve for the next order solution and so on, as described below.\footnote{There is an inconsistency in this way solving the solution in that at each order in $\delta_{L,l}$ the solution should be smoothed once by our filter, but the filter enters in a more complicated manner \citep{carroll14}.  This complication is ultimately resolved with our limiting procedure.}  Note that at lowest order $\nabla u_l$ equals $-\partial_\tau \log D(a) \delta_{L,l}$ and that $\delta_l$ equals $ \delta_{L,l}$, and so up to stochastic and higher order terms $X_\Lambda$ is proportional to $\delta_{L,l}$.
 
When computing the power spectrum with this expansion method, the ETHLSS solutions to order $\delta_l^4$ (i.e. 1-loop)
include the SPT solutions $P_{22}$ and
$P_{13}$ with $\tilde \delta_L \rightarrow \tilde \delta_{L, l}$.  One can derive these contributions to $P_{\rm EFTLSS}^{\rm 1-loop}(k)$ by taking the nonlinear terms in the eq.~(\ref{eqn:nonlinear}) evaluated with the lower order solutions and integrating them over the Green's function \cite{carrasco12}.
In addition, the other contribution at 1-loop that is unique to EFTLSS is computed in the analogous manner and given by \cite{carrasco12}
\begin{equation}
  \widetilde{[\delta c]}_{l}^{(3)} = k^2 \int_0^a da' \, G(a, a') \,
  \left(c_s^2(a') + \frac{\partial_\tau \log D(a')}{{\cal H}(a')} c_v^2(a') \right)
  \tilde \delta_{L,l} (k, a').
\end{equation}
It is useful to define
\begin{eqnarray}
\alpha_{c} &\equiv&  D(a)^{-1} \int_0^a da' \, G(a, a')
  \overbrace{\left(c_s^2(a') + \frac{\partial_t \log D(a')}{{\cal H}(a')} c_v^2(a') \right)}^{c_{\rm tot}^2(a')} D(a'), \label{eqn:alphac}\\
  &\approx&  -\frac{1}{9}  \frac{c_{\rm tot}^2(a)}{{\cal H}(a)^2}
  \quad \mbox{in the Einstein de Sitter if } c_{\rm tot}^2(a')\propto a' , \label{eqn:alphaED}
\end{eqnarray}
where the approximation comes from requiring that $\alpha_c P_{11}$ has
approximately the same $a^4$ dependence as $P_{13}$, as is required to cancel the $\Lambda$ dependence.  This was the scaling assumed in \cite{carrasco12}, but there will also be a contribution to $c_{\rm tot}$ that has a different time dependence \cite{pajer13}.

Combining all the terms, the familiar SPT terms and the new terms owing to ``sound'' and ``stochasticity'', the 1-loop power spectrum in EFTLSS is 
\begin{equation}
  P_{\rm EFTLSS}^{\rm 1-loop}(k) = P_{11, \Lambda} +
  P_{22, \Lambda} + P_{13, \Lambda} +
  2 \alpha_{c, \Lambda} k^2  P_{11, \Lambda}  + P_{J, \Lambda},
\label{eqn:1loop}
\end{equation}
where subscript $\Lambda$ is shorthand for the replacement
$P_L \rightarrow P_L \,W_\Lambda^2$ in that terms integrals or, in the case of the stochastic term, $P_J \rightarrow P_J \,W_\Lambda^2$.
The third and fourth term are the only 1-loop terms unique to EFTLSS, and we have used that $2 \langle \widetilde{[\delta c]}_{l}^{(3)} \tilde \delta_{L,l} \rangle  = 2 \alpha_{c, \Lambda} P_{11, \Lambda}$.  In contrast to previous EFTLSS studies, we have defined $\alpha_c$ to be negative for positive $c_{\rm tot}$, in part because in most 1D cosmologies we consider $c_{\rm tot}$ is negative.\footnote{The authors of \cite{baumann12} argued that the $\Lambda = \infty$ sound speed, $c_s$, is always positive and generally will dominate over the contribution from $c_v$ in 3D.  However, their argument for the positivity of $c_s$ relied on the nonlinear power spectrum growing faster than the linear one, which we find to be never the case in 1D.} 

EFTLSS assumes that modes with $k' \gg k$ do not influence $ P_{\rm EFTLSS}^{\rm 1-loop}(k) $ in eq.~(\ref{eqn:1loop}) except through the value of $2\alpha_c$.  It is easily verified that the new terms can effectively absorb the contributions that SPT receives from $k' \gg k$.  The EFTLSS ``sound'' term is proportional to $k^2 P_L(k)$, which is precisely the $k$-dependence needed to cancel the UV contribution in $P_{13}$ (i.e. to cancel the $\Lambda$ dependence).  Also, modes with $k'\gg k$ contribute a term in $P_{22}$ that scales as $k^4$ as $k\rightarrow 0$, just like the stochastic term $P_J$ (see appendix~\ref{app:stochastic}).  
  However, prior formulations of EFTLSS drop any term that has a weaker scaling in $k$ as $k\rightarrow 0$ than higher loop terms for consistency.  For the CDM case, this results in dropping the stochastic term, $P_{J}$, even at 2-loop order \cite{carrasco14}.
This may be surprising because halo models for large-scale structure explain nonlinear evolution with just a stochastic term (scaling with the abundance of halos) that goes as $k^0$ \cite{seljak00,ma00,cooray02}.  However, in power-law cosmologies  one can show that the higher order contributions to the stochastic term can be expanded as $(k/k_{\rm NL})^{4+2i}$ with order unity coefficients for integer $i\geq 0$ \citep{pajer13}.  Thus, these terms are subdominant at $k \ll k_{\rm NL}$.  In Appendix~\ref{sec:stochastic}, we show that the stochastic term is also subdominant to our 1-loop corrections in our CDM-like cosmology on scales where 1-loop perturbation theory is successful.

Once the stochastic term is dropped and $\alpha_{c}$ is determined, so that all terms in eq.~(\ref{eqn:1loop}) are calculable, the story of EFTLSS is not over.
It would be problematic for the solution to depend on $\Lambda$ on any scale at which the calculation is used.  However, for $\Lambda$ that correspond to plausible $k_{\rm NL}$, the predictions at mildly nonlinear scales of our low order EFTLSS (as formulated above) {\it do\/} tend to depend on $\Lambda$.
(For $\Lambda^{-1} = 10\,$Mpc, $[\eta^2]_\Lambda$ is suppressed by $20\%$ relative to $[\eta^2]_\infty$ while $P_{22}$ is less sensitive to $\Lambda$.)
To remove this $\Lambda$-dependence requires including terms to higher order
in $k/\Lambda$, working at lower $k$, or increasing $\Lambda$.  The approach taken in the EFTLSS literature to avoid this $\Lambda$-dependence
is to take the limit $\Lambda \rightarrow \infty$ in the standard integrals and for the EFTLSS parameters (in our case $c_{\rm tot}$).
While this is the limit of a non-perturbative field that the EFTLSS equations
were derived to avoid, the assertion of EFTLSS is that this limit makes more
sense to apply at the final step.  (The mechanics of how this limit is applied are apparent in the power-law cases discussed in section~\ref{sec:powerlaw} and suggest that this limiting procedure is not unreasonable.)  The $\Lambda \rightarrow \infty$ limit can be done by extrapolating
measurements smoothed on smaller and smaller $\Lambda$ or by using the coefficients in the $\Lambda =\infty$ expression coefficients that provide the best fit to simulations.
  For our 1D case, this extrapolation is quite simple because the sound term must add to $P_{13,\Lambda} = -k^2 [\eta^2]_\Lambda P_{L, \Lambda}$ in a manner that sums to something $\Lambda$-independent, requiring
\begin{equation}
 2 \alpha_{c, \infty} = 2 \alpha_{c, \Lambda} +
  \left([\eta^2]_\Lambda - [\eta^2]_\infty\right).
\end{equation}
In what follows, we drop the $\infty$ subscript and write $ 2 \alpha_{c, \infty} $ as  $2 \alpha_{c}$. 

In the following subsections, we develop a pseudo-Lagrangian effective theory (Section~\ref{ss:Lagrangian}) and test the predictions of EFTLSS on the nonlinear power spectrum from simulations (Section~\ref{ss:testEFTLSS}).  (Additionally, in Appendix~\ref{sec:estcs}, we use our 1D simulations to estimate $2\alpha_c$ and find consistency with the power-spectrum matching technique used in the main text.)  Foreshadowing, we find that the EFTLSS is remarkably successful.

\subsubsection{A Lagrangian effective formulation}
\label{ss:Lagrangian}
As with SPT, a deficiency of Eulerian EFTLSS is that this theory expands the (smoothed) matter displacement, $\sigma_l^2$, in density \cite{porto13}.  Since the infrared displacement is close to being non-perturbative at the BAO scale despite coming from modes where linear theory is highly applicable (which we demonstrate in detail later in section~\ref{sec:1-loop}), this leads to a large error in the predictions at this important scale.  This motivates again constructing a ``Lagrangian-space'' effective theory, which we will refer to as LEFTLSS, that resums the contribution to matter displacements from large scales, as done in \cite[][which related it to a theory of extended objects]{porto13}.   We describe a simpler but related Lagrangian approach here.

Let us image a long-wavelength and short-wavelength decomposition of the Lagrangian displacement such that $\Psi = \Psi_l + \Psi_s$.  (Somewhat confusingly, our short wavelength term $\Psi_s$ includes equally-long wavelengths as $\Psi_l$ but sourced by the coupling of small-scale, $k>\Lambda$, modes.)  We can write the short-wavelength term as an expansion in the long-wavelength fields plus a stochastic (long-wavelength uncorrelated) term
\begin{equation}
 \Psi_s \approx  2 \alpha_{c, \Lambda} \nabla \delta_l +  \nabla J,
\end{equation}
where the above expansion keeps the lowest order terms that are local in the long-wavelength fields (and, importantly, the order required to generate the 1-loop EFTLSS counter terms).  The coefficient of the former term, which encodes the long-wavelength response of couplings with small-scale modes, has been named in a familiar manner because of a later identification.  In addition, we have dropped terms that are higher order in $\delta_l$ or its derivatives, and we omit a $\nabla^2 v_l$ term as at the lowest order of interest it is degenerate with $\nabla \delta_l$.  We can now take eq.~(\ref{eqn:PkZeld}) for the power spectrum in terms of the displacement, which depends on the cumulants of $\Psi(q_1) - \Psi(q_2) $.  At our working order in $\delta_l$, it only makes sense to consider the second cumulant (remembering that in 1D LPT only the second cumulant contributes):
\begin{equation}
\langle [\Psi(q_1) - \Psi(q_2)]^2 \rangle =\langle [\Psi_l(0) - \Psi_l(q)]^2 \rangle + 2[2 \alpha_{c, \Lambda} +  4\alpha_{c, \Lambda}^2 \nabla^2][\xi_{L, \Lambda}(0) - \xi_{L, \Lambda}(q)]  + 2\nabla^2  [\xi_J(0) - \xi_J(q)], 
\end{equation}
where again $q = q_1 - q_2$ and $\xi_J$ is the correlation function of $J$.  Without worrying about whether UV sensitive terms are countered by other terms (which we consider shortly), let us take the limit $\Lambda \rightarrow \infty$.  As with this step in EFTLSS, our small scale terms do \emph{not} go to zero, such that the $\Lambda \rightarrow \infty$ power spectrum is
\begin{eqnarray}
P(k) &=& \int dq e^{- i k q} \left(e^{-k^2 \sigma_{\rm LEFT}^2/2} - 1 \right), \\
{\rm where ~~} &&\sigma_{\rm LEFT}^2 \equiv \langle [\Psi(q_1) - \Psi(q_2)]^2 \rangle \Big |_{\Lambda \rightarrow \infty}.
\end{eqnarray}
The above includes all orders in the displacement and the 1-loop order effective theory terms.  Since the aim of LEFTLSS is to keep the linear IR RMS displacement to infinite order, but then to still have consistent counter terms, it makes sense to expand the exponential in all UV sensitive terms but to keep the IR RMS displacement.  To proceed, we define the variance in the displacement contributed by modes that are less than or greater than $k_\star$ as $\sigma^2_{<}$ and $\sigma^2_{>}$, where $k_\star$ is a free parameter to which the results are hopefully relatively insensitive.  We now want to expand all the UV sensitive terms so that they only enter at their lowest (1-loop) order (i.e., keeping $\sigma_{\rm LEFT}^2$ and the $\sigma^4$ subcomponent of  $\sigma_{\rm LEFT}^4$), the order at which they are consistently countered by our terms:
\begin{equation}
P_{\rm LEFT}^{\rm 1-loop}(k) = P^{\rm ZA}_{<}(k) + k^2 \int dq \, e^{- i k q}  e^{-k^2 \sigma^2_{<}/2} \left(-\sigma^2_{>}/2  + \overbrace{k^2 \sigma^4_{>}/8}^{\{P_{22} +P_{13}\}_{>}} +2  \alpha_{c} \xi_L +  \nabla^2 \xi_J -  {\cal A}\right),
\label{eqn:PLEFT}
\end{equation}
where $P^{\rm ZA}_{<}(k)$ is the LPT power spectrum calculated using modes only with $k< k_\star$, we have dropped the $\alpha^2$ term because it is suppressed by derivatives, and $\cal{A}$ is a constant that encapsulates $\xi_L(0)$ and other terms that renormalize it\footnote{Higher order effective terms in the displacement expansion includes terms that when expanded scale in the same way.}; it is suppressed at nonzero lag by a factor of $P^{\rm ZA}_{<}(k)$ compared to other terms.
Notice that all UV divergences are now properly cancelled.    This formula is similar to the expressions derived in \cite{porto13}; see also \cite{senatore14, vlah15}.  The limit $k_\star \rightarrow 0$ with ${\cal A}=0$ of $P_{\rm LEFT}^{\rm 1-loop}$ yields $P_{\rm EFTLSS}^{\rm 1-loop}$.  

In addition, we can calculate the correlation function by Fourier transforming  eq.~(\ref{eqn:PLEFT}) in the case where $k_\star$ is not a function of $k$:
\begin{eqnarray}
\xi_{\rm LEFT}^{\rm 1-loop}(r) &=& \xi^{\rm LPT}_{<}(r)+ \frac{1}{\sqrt{2\pi \sigma_{<}^2}} \int dq \, e^{-\frac{(r-q)^2}{2\sigma_{<}^2}} \Bigg\{\frac{\sigma_{<}^2 - (r-q)^2}{\sigma_{<}^4} \left(-\sigma^2_{>}/2  + 2 \alpha_{c} \xi_L + \nabla^2 \xi_J  - \cal{A} \right) \nonumber \\
&&~~~+ \frac{3\sigma_{<}^4 - 6(r-q)^2\sigma_{<}^2 + (r-q)^4}{8\sigma_{<}^4}  \sigma^4_{>} \Bigg\}.
\end{eqnarray}
We subsequently set ${\cal A}$ to zero, but it does not have to be, and on scales where the stochastic term is white then $\nabla^2 \xi_J=0$.   We note that in the $k_\star= \infty$ case -- which we later find produces nearly identical predictions to more motivated $k_\star$ choices -- , $\xi_{\rm LEFT}^{\rm 1-loop}(r)$ and $P_{\rm LEFT}^{\rm 1-loop}(k)$ take particularly simple forms, the sum of the Zeldovich approximation prediction plus a Gaussian integral over $2\alpha \xi_L + \nabla^2 \xi_J$.

\subsubsection{Comparison of effective theories with standard approach}
\label{ss:testEFTLSS}

\begin{figure}
\begin{center}
\epsfig{file=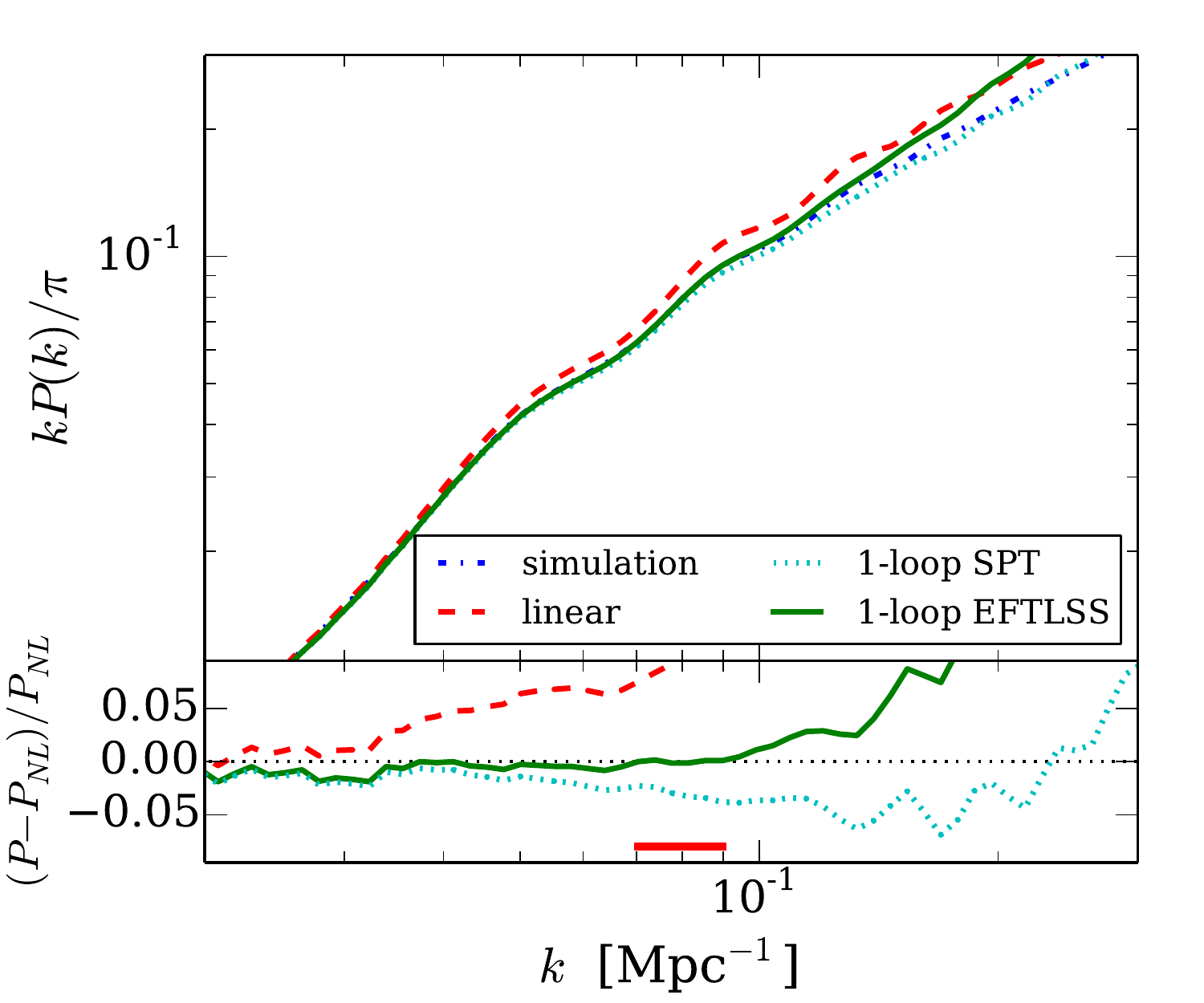, width=8cm}\epsfig{file=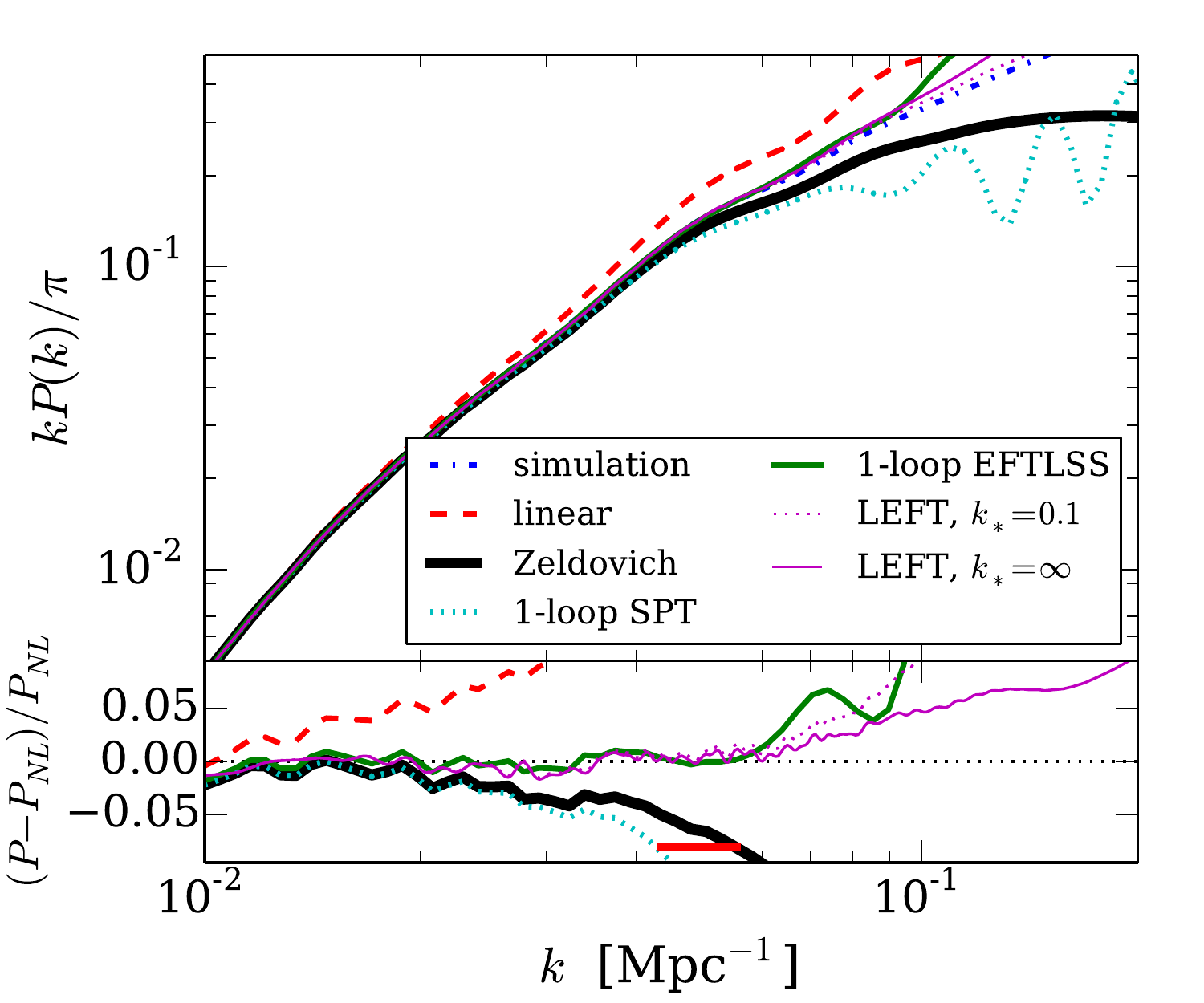, width=8cm}
\end{center}
\caption{Comparison of the power spectrum in 1-loop EFTLSS to the power-spectrum in the simulation, in Eulerian linear theory, and in 1-loop SPT.  The left panel shows $z=1$, and the right panel shows $z=0$, both in our CDM-like 1D cosmology.  The red horizontal lines span the wavenumbers over which $2\alpha_{c}$ is estimated, although it makes little difference if a factor of two smaller wavenumbers are used (as apparent from the flatness of the residuals).   In the right panel, we also show the Zeldovich approximation power spectrum and the predictions of 1-loop LEFTLSS (discussed in section~\ref{ss:Lagrangian}) where $2 \alpha_c$ is fit separately again as a free parameter.
\label{fig:pk_ETHLSS}}
\end{figure}

Figure~\ref{fig:pk_ETHLSS} shows $P(k)$ from the simulation, linear theory, 1-loop SPT, and 1-loop Eulerian effective theory (EFTLSS) for our CDM-like case at $z=0$ (right panel) and $z=1$ (left panel).   In the right panel only, we also show the Zeldovich approximation power spectrum and the predictions of 1-loop LEFTLSS.  To determine $2\alpha_{c}$, which is needed to compute the EFTLSS curves, we have fit for it using the EFTLSS formula over the range shown by the red horizontal line, rather than measure it from the small-scale behavior of the simulations.  The best-fit $2 \alpha_c$ we find to be $39\;$Mpc$^2$ for $z=0$ and to be $6\;$Mpc$^2$ at $z=1$, which we note is somewhat different than the $a^4$ scaling often assumed.  For both redshifts, EFTLSS dramatically improves the range of scales at which 1-loop perturbation theory is $1\%$ accurate:  SPT achieves this accuracy over $k < 0.05\;$Mpc$^{-1}$, whereas EFTLSS achieve it over $k< 0.1\;$Mpc$^{-1}$ for the $z=1$ case.  EFTLSS shows a comparable relative improvement over SPT at $z=0$.  Also, note the EFTLSS residuals are relatively flat over the scales where there is improved agreement:  This means that the $k^2P(k)$ scaling is correctly capturing the slope of the excess term that 1-loop SPT is missing.  It may be surprising that Zeldovich fares much worse than EFTLSS, as it is exact up to shell crossing (see right panel in figure~\ref{fig:pk_ETHLSS}). This improvement implies that EFTLSS, where the primary difference is that it fits for a term that scales $k^2P(k)$, is doing a better job at capturing the impact of regions where shells have crossed on the large-scale power.

The $1$-loop Lagrangian effective theory (LEFTLSS) curves in the right panel of figure~\ref{fig:pk_ETHLSS} are computed by dropping the stochastic $\xi_J$ contribution (as also done in EFTLSS) and by either taking $k_\star = 0.1\;$Mpc$^{-1}$ (using a Gaussian cutoff of the form $\exp[-k^2/k_\star^2]$) or $k_\star= \infty$.  The former $k_\star$ corresponds to roughly the wavenumber where the Zeldovich approximation power spectrum errs at $10\%$, and $k_\star= \infty$ corresponds to the advection owing to linear modes over all wavenumbers being resumed.  However, we find that both values of $k_\star$ yield nearly the same prediction for our CDM-like case.  For the LEFTLSS curves we again fit for $2\alpha_c$, with the fit preferring a different value than in EFTLSS of $2\alpha_c = 25~$Mpc$^2$.  That LEFTLSS is best fit with a different value is not surprising because the small-scale terms that $2\alpha_c$ is countering are somewhat different between the Eulerian and Lagrangian formulations.
  
If you squint, LEFTLSS appears to take out the BAO wiggles that are tentatively present in the EFTLSS residuals in figure~\ref{fig:pk_ETHLSS}, as others have suggested a Lagrangian theory should \cite[][see section~\ref{sec:1-loop}]{senatore14}.  In addition, LEFTLSS fares as well at predicting the power spectrum as EFTLSS.  We consider the correlation function in section~\ref{sec:1-loop}, where the improvements of LEFTLSS over Eulerian EFTLSS are even more notable.

\section{Power-law models}
\label{sec:powerlaw}

\begin{figure}
\begin{center}
\epsfig{file=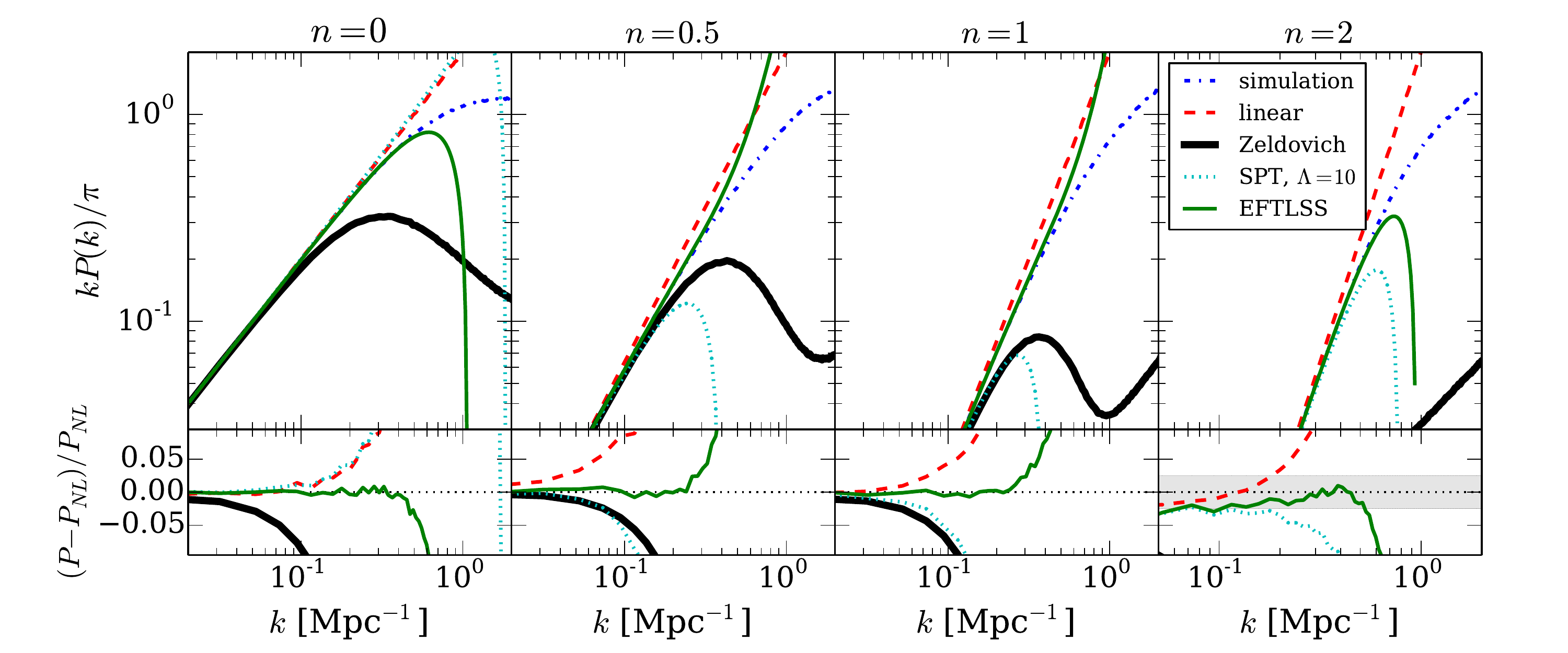, width=16cm}
\end{center}
\caption{Comparison of perturbation theories and simulations for models with
a power-law linear-theory power spectrum of the form $kP_L(k)/(2\pi) = (k/k_{\rm NL})^{n+1}$ for $k>0$, with $k_{\rm NL}=1$~Mpc$^{-1}$.  The dashed line shows $P_L$, and the dashed-dotted line shows the result of the $N$-body
simulation.  The Zeldovich approximation power is shown as the thick black solid line (computed from the modes in the simulation with $k_{\rm Nyquist}=30~$Mpc$^{-1}$), while that of 1-loop SPT is shown as the
dotted line (using a top hat $W_\Lambda$ with $\Lambda=10$).  The predictions of EFTLSS are shown as the green solid line,
where EFTLSS is the same as SPT except that we have fit an extra term that scales as $k^2 P_L(k)\sim k^2$ in the left panel, $k^2 P_L(k) \sim k^{2.5}$ in the left-middle panel,
$k^2 P_L(k) \sim k^3$ in the right-middle panel, and $P_J\sim k^2P_L(k) \sim k^4$ in the
right panel. The grey band in the rightmost case is to indicate that this particular numerical simulation is not fully converged with the estimated error being roughly half the vertical width of this band.
\label{fig:powerlaw}}
\end{figure}

Thus far we have considered our CDM-like model.
We now consider power-law power-spectrum models where the 1D linear-theory
power spectrum is 
\begin{equation}
k P_L(k)/(2\pi) = (k/k_{\rm NL})^{n+1}
\end{equation}
 for $k>0$ and the symmetric form for $k<0$.  During the Einstein-de Sitter phase, the evolution of such scale-free initial conditions is self-similar and so it is sufficient to consider a single epoch.
The panels in figure~\ref{fig:powerlaw} shows the predictions for power-law models (all
with $k_{\rm NL}=1~$Mpc$^{-1}$) with index $n=0$, $n=0.5$, $n=1$ and $n=2$, from left to right respectively.
The curves correspond to linear theory, the Zeldovich approximation
(LPT at all orders), and the fully nonlinear calculation, all computed by
taking the power spectrum of modes in the
simulation volume (i.e., using the same initial random numbers as the simulation curve; $k_{\rm Nyquist}=30~$Mpc$^{-1}$).  We have tested that the nonlinear power spectrum from the simulation (calculated using the fiducial specifications that were chosen with the CDM-like cosmology in mind) is converged for the three leftmost cases (appendix~\ref{app:PMcode}), but, even using much smaller time steps and higher initialization redshifts, we were only able to achieve convergence at the 2\% level for the rightmost $n=2$ case.\footnote{We put some effort into simulating for $n=3$, but were not able to achieve even a modicum of convergence.  Indeed, it is somewhat surprising that the $n=1$ and $n=2$ converge to a steady state (which we take to be the self-similar solution) even though the Zeldovich approximation power spectrum that is used to initialize the simulation is formally infinite for $n\geq1$ (although not our simulated case that is cutoff at  $k=10\,k_{\rm NL}$).  Because $n<-1$ cosmologies have large-scales collapse before small scales such that perturbation theory is not valid, the range of power-law cases we select essentially maximizes what is possible to simulate.} 
Figure~\ref{fig:powerlaw} also shows the predictions for $P(k)$ of an analytic calculation using 1-loop
SPT with a cutoff at $\Lambda = 10\,k_{\rm NL}$ (a cutoff is required to yield a finite answer in the rightmost two panels)
and of EFTLSS.  The $2\alpha_c$ parameter in EFTLSS has been
adjusted, by hand, to provide a good fit to the $N$-body results, and we have ignored the stochastic term as is justified for these cosmologies.\footnote{For $n=2$ the stochastic term is potentially important, but it has exactly the same form as the effective sound speed term.}    We do not show curves for LEFTLSS as this model makes less sense to apply to these power-law cases where the large-scale displacement is not necessarily the effect that is the least perturbative.  

Recall that both SPT and LPT can be expressed as expansions in $\sigma^2(q)$.
For power-law\footnote{One can find expressions for $P(k)$ for power-law models
for the full 3D case in the Zeldovich approximation in \cite{taylor96} and within SPT in
\cite{pajer13}.} models $\sigma^2(q)$ converges for $-1<n<1$,
yielding for our normalization convention
\begin{equation}
  \sigma^2(q) = 2\, k_{\rm NL}^{-n-1} q^{1-n}
                \Gamma[n-1] \sin\left[\frac{(n+2)\pi}{2}\right].
\end{equation}
Infinite order SPT and LPT are thus convergent over the same range in $n$.  That $\sigma^2$ diverges for the rightmost two panels in
figure~\ref{fig:powerlaw} perhaps explains why the Zeldovich approximation fares poorly when calculated from the modes within the box.
However, the Zeldovich approximation also deviates significantly from $P(k)$ in the simulation for the $n=0$ and $n=1/2$ cases, where $\sigma^2(q)$ is convergent.  Even Eulerian linear theory matches most of the simulations to higher $k$ than the Zeldovich approximation.  This again demonstrates that shell crossing plays an important role in the nonlinear evolution of the power spectrum, noting the exactness of the Zeldovich approximation prior to shell crossing (or of SPT to infinite order).

One convenient aspect of power-law models is that there are simple analytic forms for the 1-loop power spectrum in all the power-law cases considered in figure~\ref{fig:powerlaw}.  The 1-loop EFTLSS predictions assuming a top hat  $W_\Lambda(k)$ with a sharp cutoff at $\Lambda$ are
\begin{equation}
  P^{\rm 1-loop}_{\rm EFTLSS}(k)  =
\begin{cases}
	2 \pi + 4 \pi k \left( \log \frac{\Lambda}{\Lambda -k} - 2 \pi k^3 \frac{1}{\Lambda(\Lambda -k)}\right)  + 4\pi k^2 \alpha_{c, \Lambda} + P_{J,\Lambda}& \text{if $n=0$}, \\
	2 \pi k^{1/2} - 4\pi k^2 \frac{2 \Lambda -k}{\sqrt{\Lambda (\Lambda -k)}} + 2\pi k^{5/2} \left(2 \alpha_{c,\Lambda} - \frac{1}{\Lambda} - \frac{1}{\sqrt{\Lambda-k}}  \right) + P_{J,\Lambda} &  \text{if $n=1/2$}, \\
   2\pi k + 2\pi k^3
        \left( \log \left[(\frac{k}{\Lambda})^2(1 -{\frac{k}{\Lambda}})\right]     
  + 2\alpha_{c,\Lambda}\right) + P_{J,\Lambda} & \text{if $n=1$},\\
 2\pi k^2 +
               2\pi k^4\left[2\alpha_{c,\Lambda}- \Lambda\right]
               - \pi k^5 + P_{J,\Lambda} & \text{if $n=2$},\\
  \end{cases}
\end{equation}
 setting $k_{\rm NL}=1$ (with SPT being the limit $\Lambda\rightarrow \infty$ and $ P_{J,\Lambda}  = 0$).
Only in the $n=0$ and $n=1/2$ cases are the SPT limits of these equations convergent, reducing to $P_{\rm 1-loop}^{\rm SPT} =2\pi$ (no correction over $P_L$) and $ P_{\rm 1-loop}^{\rm SPT}  = 2 \pi k^{1/2} - 8 \pi k^2$, respectively.  In both of the $n=0$ and $n=1/2$ cases, the effective terms act to cancel the lowest order in $k/\Lambda$ dependences, and also in both of these cases EFTLSS predicts power-law scalings in $k$ than are \emph{not} present in $\Lambda \rightarrow \infty$ SPT. [SPT gives terms that scale as $k^{2n+1}$ while EFTLSS adds a term going as $k^{n+2}$ with an unknown
coefficient and $P_J$, which has a leading-order behavior of $k^4$ as $k\rightarrow 0$.]  In the $n=1$ and $n=2$ cases, the new EFTLSS terms act to exactly cancel bone fide divergences as $\Lambda \rightarrow \infty$.\footnote{The sound term also counters the $\Lambda$ dependence for the logarithmic term noting that this term is $\approx 2\log (k /\Lambda)$.}  In the $n=2$ case, both the $2\alpha_c k^2 P_L$ and $P_{J,\Lambda}$ have the same $k\rightarrow 0$ scaling ($k^4$) and so both act to cancel the SPT divergence that scales as $k^4\Lambda$.  At higher $n$, which we find are not feasible to simulate, $P_{J,\Lambda}$ rather than $2\alpha_c k^2 P_L$ cancels the lowest order divergence in SPT.  

In all power-law cases considered in figure~\ref{fig:powerlaw}, EFTLSS agrees much better with the simulations than with LPT (using the modes in the box) or with SPT (cutoff with $\Lambda=10$), providing a factor of $\approx 3$ enhancement in the scales beyond where linear theory is percent-level accurate (or the other theories for that matter). This improvement does not appear to be a coincidence because of the additional free parameter.  In the two cases where SPT yields a finite prediction, 
 EFTLSS is still a significant improvement over SPT.     While an improvement over 1-loop SPT is guaranteed because EFTLSS has a free parameter, we see that these additional terms are at least as important as the perturbation theory generated ones for $n\ge 0$, allowing us to fit the $N$-body power spectra at the percent level to about $0.3\,k_{\rm NL}$ or so  (and, actually, this is about the fraction of $k_{\rm NL}$ we found for our CDM-like case).  Fitting instead for power-law terms of the form $k^m$ with integer (or half integer) powers in $m$ and that are different than the EFTLSS terms does not work nearly as well, a result that can be gleaned by noting that the slope of the EFTLSS term is exactly what is required to correct the SPT prediction.  Pajer \& Zaldarriaga \cite{pajer13} presented a similar comparison of EFTLSS against power-law models, testing it for 3D cosmologies with $n=-1.5$ and $-1$ using data from published simulations.  They found $<5\%$ residuals, and, likely because of the accuracy of the simulations they employed, their residuals are not as striking in their flatness as those shown here.  

\section{Understanding the 1-loop behavior of these theories}
\label{sec:1-loop}

\begin{figure}
\begin{center}
\epsfig{file=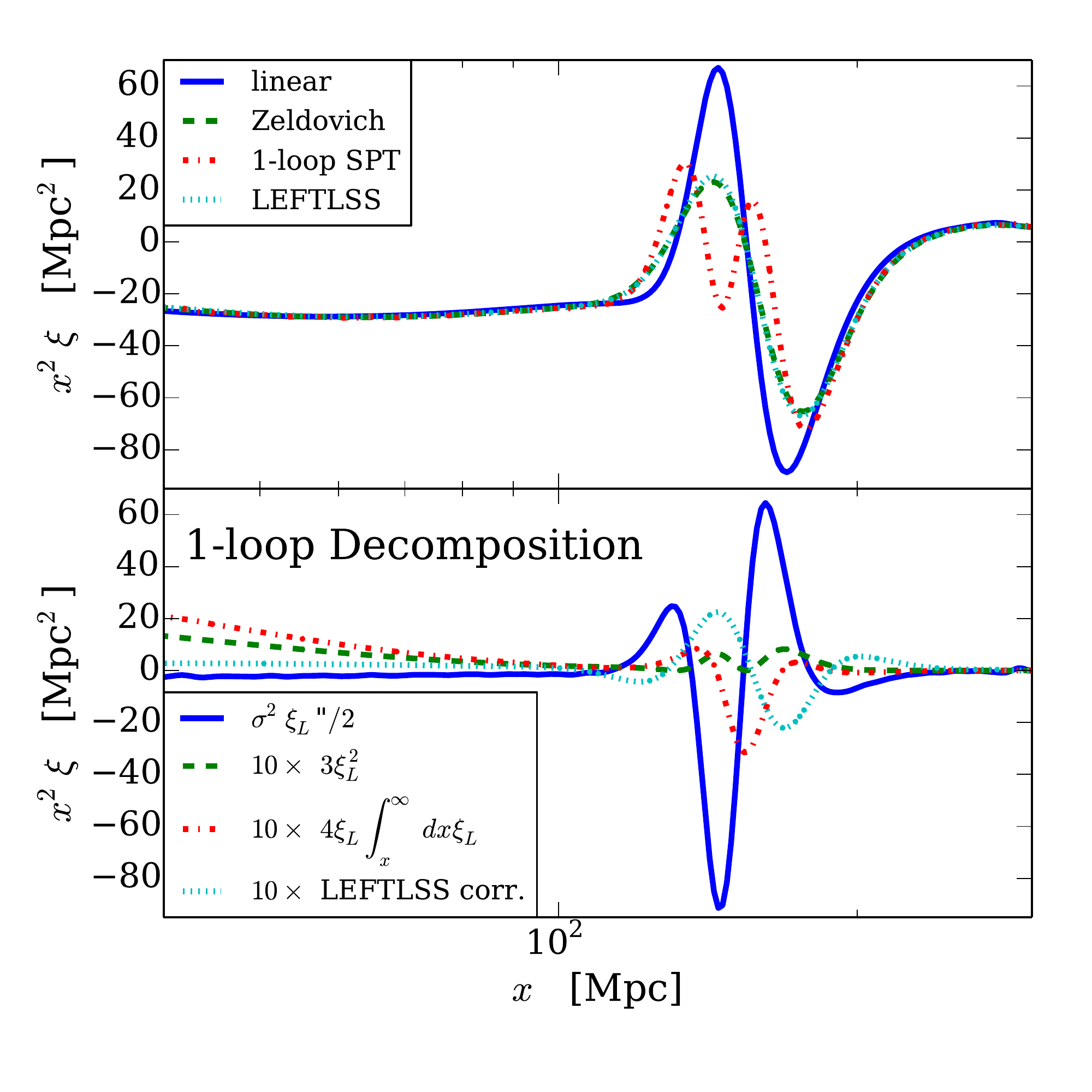, width=10cm}
\end{center}
\caption{The top panel shows the 1-loop SPT, Zeldovich approximation/LPT, and 1-loop LEFTLSS predictions for the correlation function ($\xi$), and the first three labeled curves in the bottom panel correspond to the 1-loop terms that contribute additively to $\xi$ (using eq.~\ref{eqn:xiexpan}), with the smaller terms multiplied by a factor of ten.  The other curve in the bottom panel is ten times the difference between LEFTLSS and LPT.  Note that the full nonlinear solution agrees remarkably well with the Zeldovich approximation (figure~\ref{fig:corrfuncs_comp}).    These calculations are for our 1D CDM-like cosmology and $z=1$, an epoch that shows comparable nonlinear evolution to these functions at $z=0$ in 3D.  The bottom panel shows that the RMS displacement term, $\sigma^2 \xi_L''/2$, is larger than the other 1-loop corrections by more than a factor of ten.
\label{fig:corr_decomp}}
\end{figure}

Here we attempt to understand the 1-loop predictions of these different theories and how well this order captures the actual nonlinear evolution, again specializing to the 1D case and returning to CDM-like cosmologies.  We show that this simplified case has bearing on many of the topics discussed recently in the perturbation theory literature (such as shifts in the BAO peak, the comparison of LPT to SPT, and the [dis]advantages of a Fourier space over a configuration space analysis).  

Let us first consider the predictions of these theories in configuration space.  Configuration space is the most convenient space to compare with observations of galaxy clustering.  The 1-loop correlation function is simply the Fourier transform of the 1-loop expansions for $P(k)$ (eq.~\ref{eqn:SPT1loop}; appendix~\ref{app:LPTtoSPT}):
\begin{eqnarray}
  \xi_{\rm SPT}^{\rm 1-loop}(x) &=& \xi_{L}(x) +
  \overbrace{3 \xi_L^2(x)}^{\rm growth} +
  \overbrace{4 \xi_L'(x) \int^\infty_x dx\,\xi_L(x)}^{\rm dilation} +
  \overbrace{\frac{\sigma^2(x)}{2}\xi_L''(x)}^{\rm RMS~displ.}, \label{eqn:xiexpan}
\end{eqnarray}
where primes denote derivatives with respect to the argument.  For 1-loop EFTLSS, we would replace $\sigma^2(x)$ with $\sigma_{\rm eff}^2(x) \equiv \sigma(x)^2 - 2\alpha_{c}$ in eq.~(\ref{eqn:xiexpan}).\footnote{We can write a similar expression for $\delta(x)$ to second and third order:
\begin{eqnarray}
\delta^{(2)} &=& \overbrace{\delta_L^2}^{\rm growth} +  \overbrace{ (\nabla \delta_L) \int_x^\infty dx\ \delta_L }^{\rm shift}\\
\delta^{(3)} &=& \overbrace{\delta_L^3}^{\rm growth} + \overbrace{\frac{1}{2} \nabla^2 \delta_L \left(\int_x^\infty dx\ \delta_L \right)^2}^{\rm 2^{nd}~order~shift} + \overbrace{3 \delta_L \nabla \delta_L \int_x^\infty dx\ \delta_L }^{\rm growth-shift}.  
\end{eqnarray}
Somewhat confusingly, both shift and growth terms in $\delta$ can mix to generate what we call a, e.g., growth term in $\xi(x)$ (eq.~\ref{eqn:xiexpan}).}
The $2\alpha_{c}$ term is a $15\%$ correction to $\sigma^2$ in
our CDM-like cosmology, and we shall ignore this correction in most of what
follows as this difference has little impact on our discussion.

\begin{figure}
\begin{center}
\epsfig{file=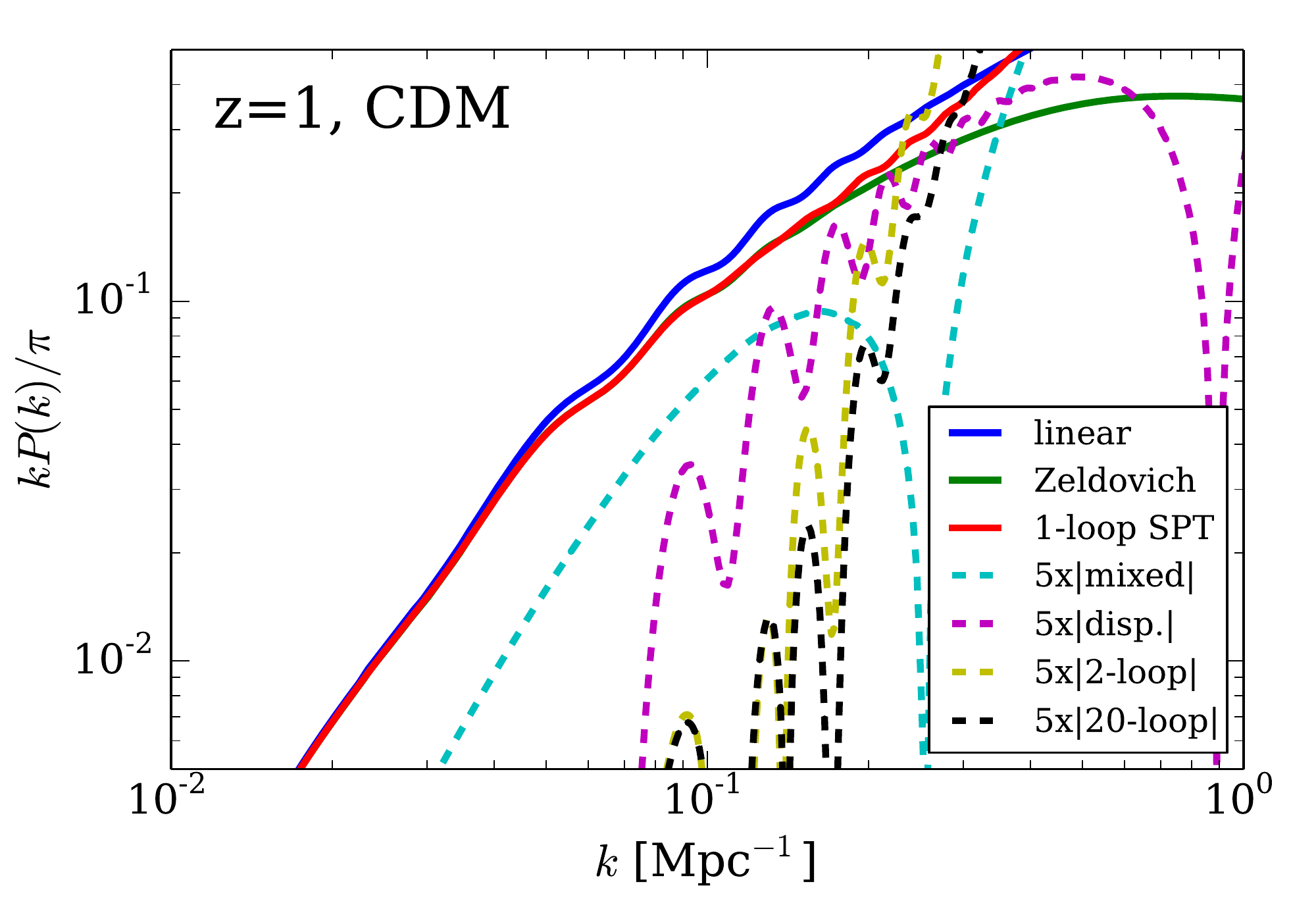, width=10cm}
\end{center}
\caption{Decomposition of our CDM-like 1D power spectrum at $z=1$.  The labels distinguish terms that owe to the RMS displacement from ``mixed'' terms that mostly arise from other effects (see eq.~\ref{eqn:PSPTdecomp}).  Also shown is the absolute difference between  the 2-loop and 1-loop power spectrum and the same but between the 20-loop and the 1-loop, both multiplied by $5$ (labelled $5\times|2$-loop$|$ and $5\times|20$-loop$|$).  The RMS displacement effect is comparable to other 1-loop terms for the first and second BAO peaks in the power spectrum, in contrast to the correlation function, and the SPT expansion also converges more quickly at these scales than at the BAO peak in the correlation function.
\label{fig:pk_decomp}}
\end{figure}

Each of the terms in eq.~(\ref{eqn:xiexpan}) can be thought of as the lowest order expansion with respect to a certain effect.  The first term rightward of the linear term owes to the volumetric average of how regions grow relative to linear theory, the second term encodes the
dilation that large-scale fluctuations impart on smaller separation correlations,\footnote{In overdense regions, the large-scale overdensity acts like a closed universe remapping $r$ to slightly smaller scale.  This effect does not cancel when averaging over underdense and overdense regions because there is more growth in overdense regions, such that they are weighted more heavily than underdense ones (as we show shortly).} and the final term is the effect of the RMS matter displacement that owes primarily to modes with wavelength $\gtrsim x$.
The bottom panel in figure~\ref{fig:corr_decomp} shows the contribution of each of these terms to $x^2 \xi$, a quantity that emphasizes the BAO scale.  The solid curve in the bottom panel represents the dominant correction owing to the RMS displacement, $\sigma^2 \xi_L''/2$. The other terms that appear in eq.~(\ref{eqn:xiexpan}) [all multiplied by a factor of ten in figure~\ref{fig:corr_decomp}] result in much smaller corrections at the BAO scale and indeed even smaller corrections to the underlying continuum at $x \gtrsim 10\,$Mpc for the CDM-like case at hand (all 1-loop corrections scale with time in the same way).
This result is not surprising when looking at these terms in detail.
In CDM, the RMS displacement at the BAO scale is
$\sigma (100 \, \Mpc) \simeq 10\,$Mpc, comparable to the width of the BAO
feature and thus to $1/\sqrt{\xi_L''}$.
Hence, the RMS displacement term, $\sigma^2 \xi_L''/2$, is an ${\cal O}(1)$
correction.
The growth term, $3\xi_L^2$, is small because $|\xi_L(x)| \sim 10^{-3}$ at
$x\sim 10^2\,$Mpc and $z=1$ (see figure~\ref{fig:pk_decomp}).
The dilation term, $4x \xi_L'\langle \delta_L^2\rangle_x$, is also small because
$\langle\delta_L^2\rangle_x\equiv x^{-1}\int_x^\infty dx\,\xi_L\sim -10^{-3}$
and $x \xi_L'\sim 10\,\xi_L$ at the BAO peak.
The same result that the RMS matter displacement is the main source of
evolution of the BAO feature in the correlation function applies to the
full 3D CDM case \cite{tassev14a, senatore14}.

Despite its smallness, the dilation term has received some attention because it can shift the peak position of the baryon acoustic scale \cite{CroSco08,PadWhi09,sherwin12}, whereas the other nonlinear terms do not result in a significant shift because they are quadratic in derivatives of $\xi_L$ or, in the case of $\sigma$, have a broadband response.  To compute the BAO peak shift from the dilation term we note that $\xi_L(x+\epsilon \, x) \approx \xi_L(x) + \epsilon x \xi_L'(x)$.  Thus, the dilation term results
in a displacement in the BAO peak in the correlation function of $\epsilon \approx 4 \langle \delta_L^2(x\sim 100 \Mpc)\rangle$.
The 1D shift using 1-loop SPT is $3.2$ times larger than the shift of
$68/63\,\langle \delta_L^2(r\sim 100 \Mpc)\rangle$ found in \cite{sherwin12} for the full 3D case.  The authors of \cite{sherwin12} showed that the 3D case can be understood from the following factors: that the local correlation function is enhanced at lowest order by a large-scale overdensity, $\delta_{\rm LS}$, by $(1+ \frac{34}{21} \delta_{\rm LS})^2 \approx 1+ \frac{68}{21} \delta_{\rm LS}$, section~\ref{sec:SPT}, and the correlation function is shifted by large-scale overdensities as $\approx \xi_L( [1 +\delta_{\rm LS}/3] \bfr)$ owing to comoving scales being contracted by the large-scale overdensity.  Thus, in 3D, expanding $\xi_L$ around $\bfr$, 
\begin{equation}
\xi(\bfr) \approx \left \langle \left(1+ \frac{68}{21} \delta_{\rm LS}\right)\left(\xi_L(\bfr) + \frac{\delta_{\rm LS}}{3} \bfr \cdot \nabla \xi_L(\bfr) \right) \right \rangle ~ \approx \xi_L(\bfr) + \frac{68}{63} \langle \delta_{\rm LS}^2 \rangle \; \bfr \cdot \nabla \xi_L(\bfr).  \nonumber
\end{equation}
Repeating the same exercise in 1D, the correlation function is modulated locally by $(1 + 2\delta_{\rm LS})^2$ (section~\ref{sec:SPT}) and the 1D shift is $\xi_L( [1 +\delta_{\rm LS}] x)$.  Combining these effects in the same manner yields $4\,\langle \delta_{\rm LS}^2\rangle\,x\,\xi_L'(x)$, exactly our 1D dilation term.

Even at $z=1$ in our fiducial CDM-like cosmology,  1-loop SPT fares poorly at describing the shape of the BAO in
the correlation function (top panel; figure~\ref{fig:corr_decomp}).\footnote{Even though 1-loop is not terribly successful at capturing the nonlinear evolution at BAO scales in our 1D CDM-like example, we note that the failure is even worse in 3D CDM  where the 1-loop prediction in the correlation function is infinite!}  The lack of success owes to 1-loop order dropping terms that are higher order expansions in the RMS displacement, of the form $\sigma^{2m} [\nabla_x^2]^{m}\xi_L(x)$.  We have checked that the term $\sigma^4 \xi_L''''(r)/8$, which enters at $2$-loop order, is not substantially smaller than the dominant nonlinear term in the correlation function $\sigma^4 \xi_L''''(r)/2$ at $z=1$, and it grows in importance relative to lower loop terms with time.  This demonstrates that SPT is only going to be weakly convergent in the correlation function at BAO scales.  In contrast, LPT includes all orders of these RMS-displacement terms (in both 1D and 3D) and hence should fair much better.  This inclusion was also the motivation for LEFTLSS.  Figure~\ref{fig:corr_decomp} shows the 1-loop LEFTLSS calculation using our best-fit EFTLSS value of $\alpha = 6~$Mpc$^2$ and $k_*=\infty$ (section~\ref{ss:testEFTLSS}; although we find using $k_*= 0.1~$Mpc$^{-1}$ is essentially indistinguishable).\footnote{Really we should fit for this parameter in LEFTLSS, but this matters little here.  At $z=0$, such fitting resulted in a $30\%$ smaller $\alpha_c$.}  The LEFTLSS prediction is nearly identical to the very successful Zeldovich approximation one, further confirming that the UV-sensitive corrections are small at BAO scales.\\

Now we turn to Fourier space and the 1-loop corrections to the matter power spectrum.
The density, dilation, and RMS displacement terms that constitute the 1-loop corrections in the correlation function (eq.~\ref{eqn:xiexpan}) each have a
power spectrum that is generally much greater in absolute magnitude than
the linear-theory power spectrum at all wavenumbers; there are large cancelations when they are summed into the 1-loop SPT power spectrum.
However, we can rewrite the terms in a manner that removes the terms that cancel, decomposing $P_{\rm SPT}^{\rm 1-loop}$ as a term that mixes
the different effects and a term that derives purely from the RMS displacement:
\begin{equation}
  P_{\rm SPT}^{\rm 1-loop}(k) =  P_L(k) +
  \int_{-\infty}^{\infty} \frac{dk'}{2\pi}
  \left\{ \overbrace{2 \frac{k}{k'}P_L(k') P_L(k-k')}^{\rm mixed} +
  \overbrace{\frac{k^2}{k'^{2}}P_L(k') P_L(k-k') -
  k^2 \eta^2 P(k) \delta^D(k') }^{\rm RMS ~displ.} \right \},
  \label{eqn:PSPTdecomp}
\end{equation}
where we could replace $\eta^2$ with $\eta^2 - 2\alpha_{c,\infty}$ to yield EFTLSS,
but we shall neglect this small change
in what follows.
Figure~\ref{fig:pk_decomp} plots the mixed and RMS displacement terms at $z=1$
in our CDM-like case.
The mixed term is larger than our problematic term for the first two BAO peaks
(the peaks where SPT is used in 3D), and unlike the RMS term does not show acoustic
oscillations.
Also shown are the contribution of higher loops.  In particular, the yellow dashed curve is the absolute difference between the 2-loop and 1-loop power spectrum multiplied by $5$, and note that the terms at each loop order grow as $a^{2(n-1)}$ and, hence, would be more important at $z=0$.  The black dashed curve is the same but between the 20-loop power spectrum and the 1-loop. The beyond $1$-loop contributions are smaller at the first couple peaks (and including 19 more loops makes little difference at $k < 0.13~$Mpc$^{-1}$ for $z=1$), suggesting that Eulerian formulation of perturbation theory (SPT and EFTLSS) yields a convergent solution for the first $\sim 2$ peaks in the case at hand.  (Of course for 1D SPT, the convergence is to the Zeldovich approximation solution and not the true nonlinear solution.)  The lack of convergence in the correlation function at the BAO peak -- discussed in the first half of this section -- owes to contributions from shorter wavelengths at which SPT is not trusted in Fourier space.  This 1D picture differs somewhat from that in \cite[][albeit in 3D]{tassev14a}, who argued that the agreement at BAO scales between SPT and the full nonlinear solution is spurious, appearing better than it actually is because the continuum power (for which SPT is more justified) is comparable to the power in BAO.  Lastly, note that the contribution from $>1$-loop terms show significant oscillations around the BAO, arising from the RMS displacement expansion.  As in the correlation function, in the power spectrum the displacement terms are the ones that converge slowest.

\section{Conclusions}
\label{sec:conclusions}

This paper discussed cosmological perturbation theory in the specialized case of 1D dynamics (e.g.~gravity between expanding sheets).  In 1D, linear-order Lagrangian perturbation theory (LPT) is exact up to shell crossing and all nonlinear terms in LPT are zero.   We further showed that infinite-order standard Eulerian perturbation theory converges to linear-order Lagrangian perturbation theory (and is easily calculable for any given order). Unfortunately, for a wide variety of initial conditions, we found that these standard perturbation theories do not converge to the nonlinear solution at any mildly nonlinear scale, indicating that dynamical equations that allow for shell crossing are needed.

In this spirit, we considered the recently developed effective field theory of large-scale structure (EFTLSS), both the standard Eulerian space formulation and a Lagrangian space derivative.  This theory attempts to more rigorously formulate perturbation theory by using dynamical equations that depend only on smoothed fields whose RMS is $<1$.  
  When the effective sound speed that appears in EFTLSS was treated as a free parameter, we found that at 1-loop order EFTLSS extended the range over which perturbation theory is percent-level accurate by a factor of two in wavenumber in a CDM-like cosmology in which the overdensity has the same variance per $k$ as the concordance cosmology (consistent with the findings of 3D studies; e.g. \cite{carrasco12, carrasco14}).  We also tested EFTLSS in a large range of power-law 1D cosmologies, where we found its improvements over previous theories were even more dramatic.  The new terms that EFTLSS predicts based on symmetry are exactly those required to correct previous theories.    

Specializing to 1D also allowed us to break up the lowest order nonlinear contributions to the matter power spectrum into components that correspond to three distinct physical effects:  the RMS displacement of particles, the effect of an overdensity on its own growth, and the modulation of smaller structures by large-scale ones.  At the BAO scale in the correlation function for a CDM-like case, this allowed us to show that the primary source of nonlinear evolution is the large-scale RMS displacement that LPT captures exactly but that are treated perturbatively in SPT and Eulerian EFTLSS.  In a CDM-like cosmology, we showed that the 1-loop expressions for these theories provides a poor description for the correlation function at the BAO peak owing to the largeness of the RMS displacement, which the 1-loop theory has expanded in despite the expansion converging slowly at such scales.  This echoes similar findings in 3D, but we highlight that the analytics in 1D are substantially simplified.  In contrast to the case of BAO-scales in the correlation function, in the power spectrum the RMS displacement term is comparable to other nonlinear terms for the first few acoustic peaks.  We also discussed beat coupling and the related shift of the BAO peak by large-scale couplings in our 1D case.

Of course, the ultimate aim of perturbation theory studies is a rigorous understanding of these theories in 3D.  However, our investigation of the dynamics specializing to 1D has several advantages, including (1) much smaller simulations are required to estimate the nonlinear solution to the required precision, (2) the solutions for standard perturbation theories can be computed to arbitrary order, and (3) the derivations are considerable simplified.  Furthermore, the analogies between effects in this highly symmetric case and the full 3D one are truly striking, making the 1D case an excellent means of understanding the complex phenomena involved in the formation of large-scale structure.\\

\noindent \emph{Acknowledgements}: We would like to thank Anson D'Aloisio, Uros Seljak, and Leonardo Senatore for useful discussions, and we thank Uros Seljak and Matias Zaldarriaga for comments on an earlier version of the manuscript.  MM acknowledges support from the National Aeronautics and Space Administration through the Hubble Postdoctoral Fellowship and also from NSF grant AST 1312724.  This research was supported in part by
the National Science Foundation through XSEDE resources
provided by the San Diego Supercomputing Center
(SDSC) and through award number TG-AST120066.  We thank the Aspen Center for Physics (NSF grant
\#1066293) for their hospitality during which this work was completed.

\appendix

\section{Particle-mesh code and convergence}
\label{app:PMcode}

While 1D gravity can in principle be integrated exactly as the forces are
determined simply by the ordering of mass elements \cite{yano98, schulz13},
we have opted for the simple and fast particle mesh (PM) method.
The PM method has been used extensively for cosmological $N$-body simulations
in part because force computations boil down to Fast Fourier Transforms
\cite{hockney81, efstathiou95}.
In the PM method, forces are computed on a grid, and these forces are used to move particles.  We have written a 1D PM code, as also done in \cite{melott83, miller10}, that employs the $2^{\rm nd}$ order-accurate leap-frog integrator with time steps that are uniform in $\sqrt{a}$, with this choice chosen by trial and error from amongst steps with the parameterization $a^p$ for real $p$.  Our code uses CIC interpolation to both project the mass elements onto a grid and to evaluate forces from a grid of $\nabla \phi$.  In addition, we also smooth the force by an additional CIC kernel such that the gravitational potential is damped in Fourier space by $\sin^2(k \Delta x/2)/(k \Delta x/2)^2$ to reduce particle noise on the cell scale, where $\Delta x$ is the size of a mesh element, as motivated in \cite{hockney81}.  The initial conditions use the Zeldovich approximation displacements.  Because the
Zeldovich approximation is exact until shell crossing in 1D, this allows us to start our 1D simulations at later times than
typically done in 3D simulations.  Our PM calculations start at $z=10$ unless otherwise specified.

There are a couple confusing conceptual issues for 1D cosmological simulations.  First, what does it mean to compute forces from an infinite series of sheets (or, in our case, a periodic domain with $N$ sheets)?  The force on any sheet is proportional to the number of sheets to the left minus the number of sheets to the right (both infinite).  In a periodic box of length $L$, the force at $x$ is the number of sheets from $(x, L)$ minus that from $(x, 0)$.  Thus, every sheet in the box feels a force except the center sheet.  At first glance this violates translational invariance, but note that the Hubble friction exactly counters this gravitational force.  If sheets have not crossed, how do density structures develop as there is no net force?  The answer is that sheets start off with different peculiar velocities (as determined by the $\delta_L$) and, hence, there will be convergences and divergences that affect $\delta$.  A final point of clarity is that the sheets must have constant surface density in comoving space in order to conserve mass.  

To simulate the dynamics of $10^{8}$ sheets with $10^7$ PM elements from
$z=10$ to $z=0$ with our code requires about an hour on one CPU.  We have parallelized our code over shared memory, which has enabled us to run simulations that have up to a billion elements (with the limiting factor from running larger runs being memory; all the simulations reported here use $<64~$Gb).  Our typical simulation involves  $10^{8}$ sheets with $10^8$ PM elements,  which is more than sufficient for the simulation to estimate the nonlinear $P(k)$ to $1\%$ accuracy.  In fact, the smaller number of resolution elements needed is a significant advantage of 1D studies:  In 1D, a simulation volume of
\begin{equation}
  V_{\rm 1D} = 1.3\times 10^7 ~{\rm Mpc}~~ f_{1\%}^{-2}
    \left( \frac{0.01 \;{\rm Mpc}^{-1}}{\Delta k} \right)
\end{equation}
is required to be able to measure the power spectrum with accuracy of $f_{1\%}$ (where $f_{1\%}$ is in units of 1\%),
assuming Gaussian statistics and a bin width in $k$-space of $\Delta k$.  In contrast, in 3D a volume of 
$k=|\bfk|$ requires
\begin{equation}
  V_{\rm 3D} = 4 \times 10^9~ {\rm Mpc}^3~~ f_{1\%}^{-2}
  \left(\frac{0.1 \;{\rm Mpc}^{-1}}{k}\right)^2
  \left(\frac{0.01 \; {\rm Mpc}^{-1}}{\Delta k} \right),
\end{equation}
assuming all wavenumbers that fall within a shell of width $\Delta k$ are binned to measure the power spectrum. Even if convergence is reached with merely $1$ particle per Mpc$^n$-- a typical resolution of 3D simulations aimed at precisely measuring the evolution at mildly nonlinear scales \cite{carlson09} and what we find here is needed in 1D --, $4\times10^{9}$
particles are required in dimension $n=3$ for $k=0.1\,$Mpc$^{-1}$ and $\Delta k =0.01\,$Mpc$^{-1}$ versus just $1\times10^{7}$ in $n=1$ at any $k$ and the same $\Delta k$ to reach $1\%$ fractional precision. 

\begin{figure}
\begin{center}
\epsfig{file=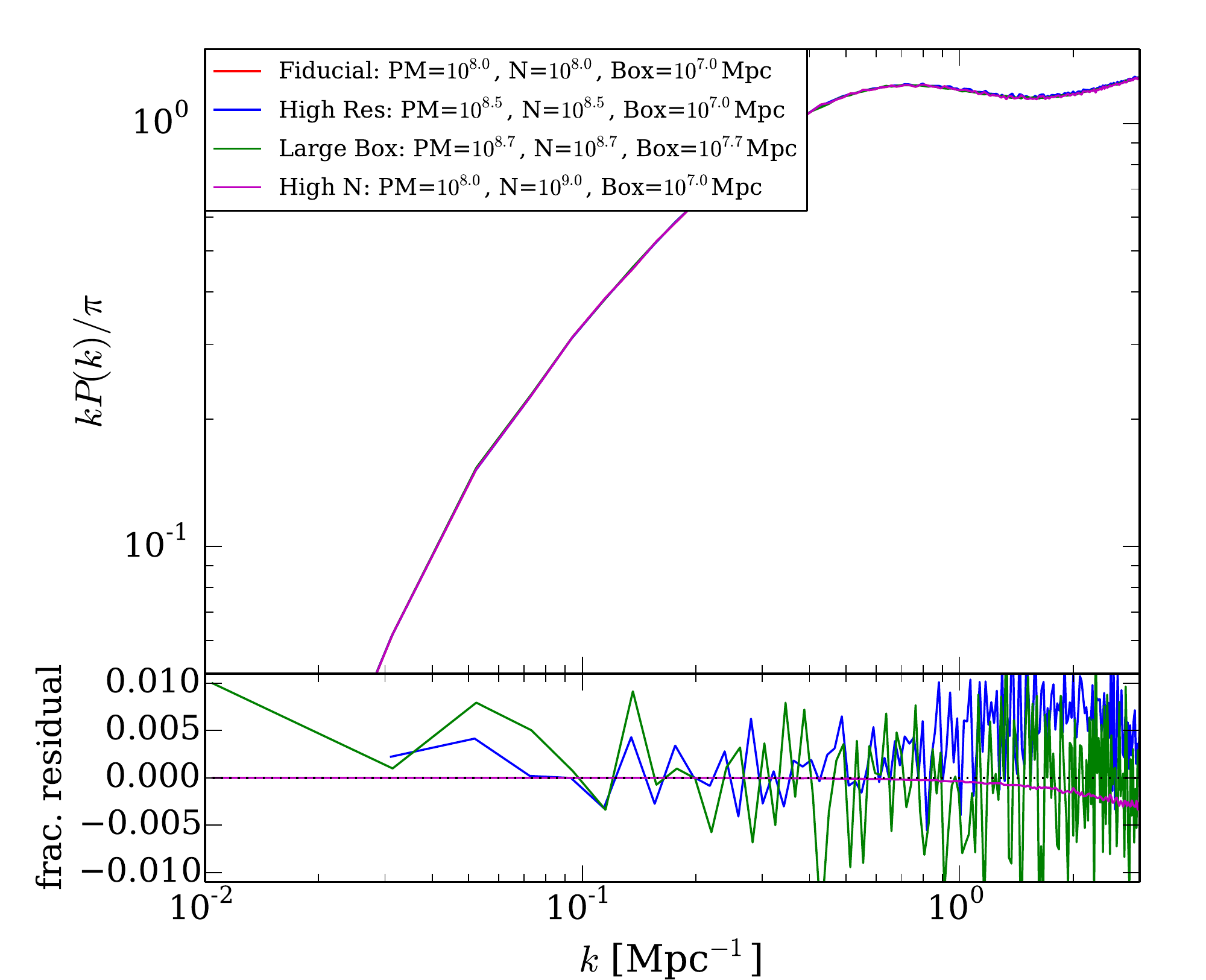, width=7.5cm}
\epsfig{file=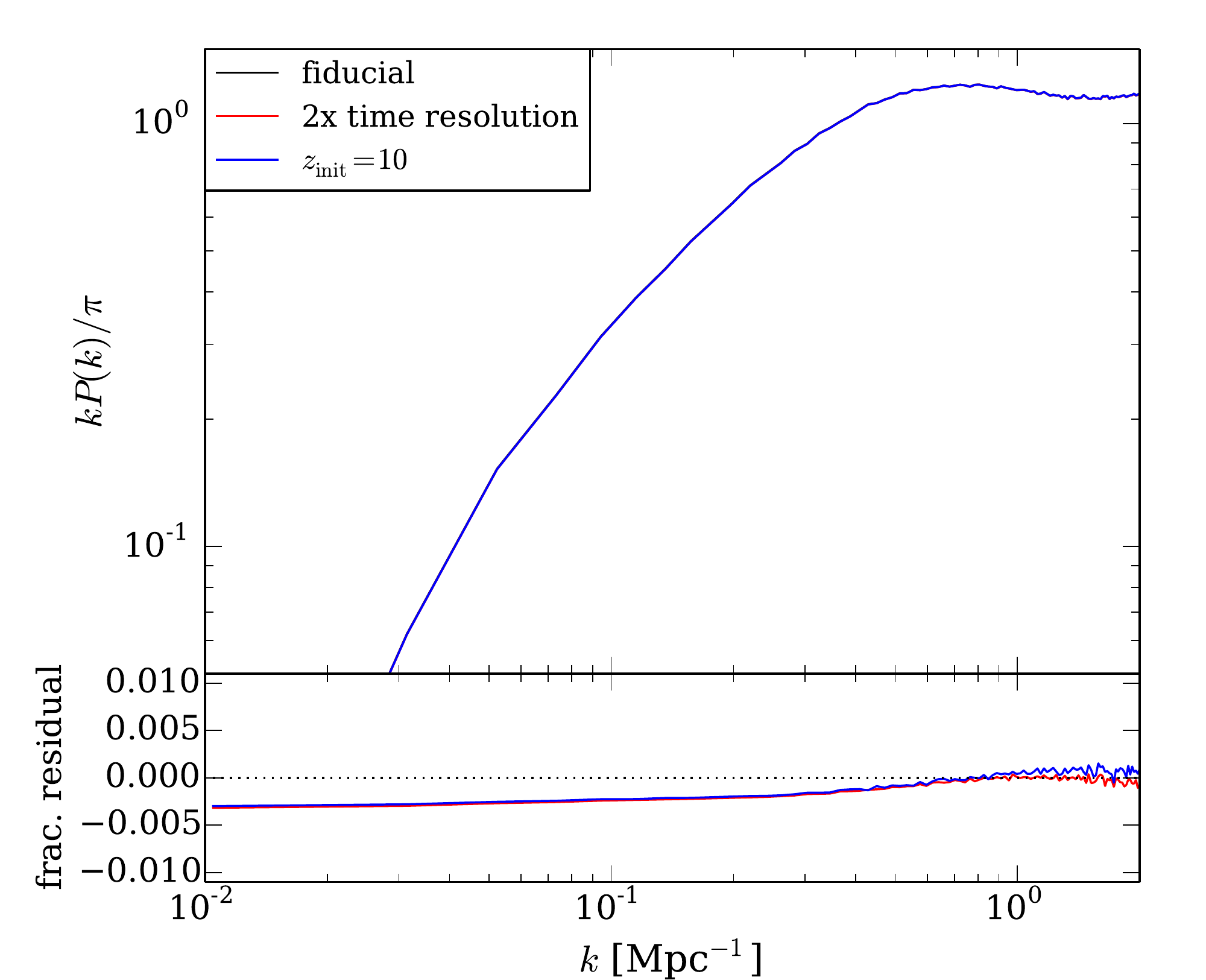, width=7.5cm}
\end{center}
\caption{Convergence tests varying the PM grid size, the number of sheets, the box size, the initialization redshift, and the time step for the $z=0$ CDM-like power spectrum.  Models with fixed box size have the same random numbers for all overlapping modes.  For all cases we achieve $<1\%$ precision.
\label{eqn:convert}}
\end{figure}

Figure~\ref{eqn:convert} shows convergence tests for the $z=0$ CDM-like case.  The lefthand panels show the convergence in PM grid size (denoted ``PM''), number of mass elements (``N''), box size (``Box''), where each estimate of $P(k)$ has used $30,000$ modes.  Runs with the same box size have the same random numbers for overlapping modes.  We find that the most important variable is resolution, with convergence at $k < 1~$Mpc$^{-1}$ at the percent level once the PM resolution is $\leq 1~$Mpc and the mean inter-sheet spacing satisfies the same inequality.  In the main body of the paper, we use a fiducial resolution of $0.1~$Mpc,  box sizes $\geq 10^7~$Mpc, and number of particles and PM grids $\geq 10^8~$Mpc.  The inequalities arise because in the CDM-like case our investigations benefit from even higher precision than $\sim 1\%$ and so we run the largest simulation possible (for the power-law cases equality holds).  The righthand panel in figure~\ref{eqn:convert} compares the power spectrum in simulations that change the initialization redshift and time stepping. For the fiducial choices of $z=20$ and $\Delta \sqrt{a} = 0.001$ the calculations are converged to a few tenths of a percent level in both parameters.  (That the lower initialization redshift simulation has similar residuals to the small time step simulation suggests that much of the error occurs at the highest redshifts, likely indicating that our time stepping criterion becomes too coarse there.)  We conclude that our code is converged to sub-percent for the desired calculations in the CDM-like cosmology.  
  
  We have run separate convergence tests for the power-law cosmologies discussed in section \ref{sec:powerlaw} that vary the initialization redshift (but keep $0.1$~Mpc resolution and the fiducial time step).  For these calculations we maintain the box sizes and resolutions indicated above.  Only for the $n=2$ power-law case do we have to increase the initialization redshift to $z=40$ in order to achieve the desired level of convergence.

\section{The relationship between SPT and LPT}

\subsection{Proof that 1D LPT is identical to 1D SPT at every order in $\delta_L$}
\label{ap:proof}

We first derive a useful recurrence relation obeyed by
both SPT and LPT.  Starting with the SPT relations
(copying eq.~\ref{eqn:SPT_rec}):
\begin{equation}
F_n = (2n+1) X_n + Y_n; ~~~~G_n = 3 X_n + n Y_n, \label{eqn:SPT_rec2}
\end{equation}
and solving for $Y_n$ using the first equation and, then, substituting this
relation into the second equation yields
\begin{eqnarray}
G_n &=& 3 X_n + n \left[ F_n- (2n+1) X_n \right],\\
	&= & n F_n - (2n+3)(n-1) X_n,\\
         &=& n F_n - \sum_{m=1}^{n-1} \frac{k}{K_1} G_{m} F_{n-m} ,
\label{eqn:SPTrecurrence}
\end{eqnarray}
where the last line uses the definition of $X_n$ (eq.~\ref{eqn:Xn}).

We now show that, after symmetrizing, the SPT relation given by eq.~(\ref{eqn:SPTrecurrence}) is identical
to the symmetrized LPT recursion relation.    Our starting point is to derive an equation that relates
velocities and overdensities in LPT (which turns out to be the same recurrence relation as just derived, eq.~\ref{eqn:SPTrecurrence}).  This relation will allow us to derive $G_n$ from
the $F_n$ of the Zeldovich approximation and then to show that the two satisfy
both SPT recursion relations.
We can relate the two using the following equation for the LPT momentum field:
\begin{eqnarray}
  (1+ \delta_{ZA}) u_{ZA} & = &  \int d q \; \dot \Psi(q)\, \delta^D [x - q - \Psi_{\rm ZA}(q)].
\end{eqnarray}  
Going to Fourier space and taking the gradient of the above yields the relation
 \begin{eqnarray} 
 \widetilde{\theta_{ZA}} +  \widetilde \delta_{ZA} \star \widetilde{ \theta_{ZA}}  + \widetilde{\nabla \delta_{ZA} } \star \widetilde{u_{ZA}}
    & = & \int dq\,e^{-ikq} \left[-ik \dot \Psi_{\rm ZA}(q) \right] \sum_{n=0}^{\infty}\frac{[-ik\Psi_{\rm ZA}(q)]^n}{n!}, \label{eqn:fouriermomentumeqn}\\
                                        	&=& \sum_{n=0}^\infty  (n+1) H  \int \frac{dk_1 \cdots dk_{n+1}}{(2\pi)^{n}}  \delta^D \left(\sum k_i -k  \right)  F_{n+1} . \nonumber 
\end{eqnarray}
Identifying terms in this equation that have the same power in $\delta_L$ yields
\begin{equation}
G_n+ \overbrace{ \sum_{m=1}^{n-1}G_{m} F_{n-m} }^{\widetilde{\delta_{ZA}} \star \widetilde{\theta_{ZA}}} + \overbrace{ \sum_{m=1}^{n-1} \frac{k-K_1}{K_1}G_{m} F_{n-m}}^{\widetilde{\nabla\delta_{ZA} } \star \widetilde{u_{ZA}}}  = n F_n, \label{eqn:firstFnGnrecLPT}
\end{equation}
where we have labeled the less-trivial convolution terms, and are using the shorthand $G_n = G_n(k_1, \cdots, k_n)$ and $F_{n-m}  = F_n(k_n, \cdots, k_m)$.  Simplifying eq.~(\ref{eqn:firstFnGnrecLPT}) gives
\begin{equation}
G_n+ \sum_{m=1}^{n-1}\frac{k}{K_1} G_{m} F_{n-m}  = n F_n, \label{rec:LPT}
\end{equation}
and we can easily symmetrize this relation by averaging each term by taking all permutations of the $k_i$.
We will denote this symmetrizing operation on term $X$ as
$X^{\rm sym} = {\rm Sym}[X]$.
Note that the symmetrized version of eq.~(\ref{eqn:firstFnGnrecLPT}) is the same relation as satisfied by SPT (eq.~\ref{eqn:SPTrecurrence}).  Even though in both LPT and SPT we derived this recursion relation using our 1D expressions, this derivation generalizes to 3D and shows that SPT and LPT share at least one recursion relation.  Physically this relation owes to the continuity of $u$ and $\delta$.\\

We now show that $G^{\rm sym}_n = F^{\rm sym}_n$ satisfies recursion relation~(\ref{rec:LPT}), using our result that $F^{\rm sym}_n =  \frac{1}{n!}\ \frac{k^n}{\prod k_i}$ (eq.~\ref{eqn:FnLPT}).  For this to hold, eq.~(\ref{rec:LPT}) reduces to
\begin{eqnarray}
(n-1) F^{\rm sym}_n &\stackrel{?}{=}& \sum_{m=1}^{n-1} {\rm Sym}[ \frac{k} {K_1} F^{\rm sym}_{m} F^{\rm sym}_{n-m} ];\\
\frac{(n-1) k^n}{n! \prod_{i=1}^n k_i} &\stackrel{?}{=}& \sum_{m=1}^{n-1} {\rm Sym}[\frac{k K_1^{m-1} K_2^{n-m}  }{ m!(\prod_{i=1}^m k_i) \;(n-m)!(\prod_{i=m+1}^n k_i)} ];\\
(n-1) k^n &\stackrel{?}{=}& \sum_{m=1}^{n-1} \frac{n!}{m!(n-m)!} {\rm Sym}[k K_1^{m-1} K_2^{n-m}   ]; \\
&\stackrel{?}{=}& \sum_{m=1}^{n-1} \frac{n!}{m!(n-m)!}  k \; {\rm Sym}[{\rm Sym}[K_1]^{m-1} {\rm Sym}[ K_2]^{n-m}];  \\
&\stackrel{?}{=}& k^n \sum_{m=1}^{n-1} \frac{n!}{m!(n-m)!}  \frac{m^{m-1} (n-m)^{n-m}}{n^{n-1}}; \label{eqn:sym} \\
n-1 &\stackrel{\checkmark}{=}& \sum_{m=1}^{n-1} \frac{n!}{m!(n-m)!}  \frac{m^{m-1} (n-m)^{n-m}}{n^{n-1}},  \label{eq:identity} \\
\end{eqnarray}
where to reach eq.~(\ref{eqn:sym}) we have used that ${\rm Sym}[K_1] = (m/n) k$ and ${\rm Sym}[K_2] = ([n-m]/n) k$.  The final line in this reduction (eq.~\ref{eq:identity}) is an identity (which we have verified up to $n=1000$ using {\it Mathematica} and which holds in the $n\gg 1$ Sterling's approximation limit).

Now that we have $F^{\rm sym}_n$ and $G^{\rm sym}_n$ in 1D LPT, it is simple to show that they also satisfy the SPT recursion relations.  Since
\begin{equation}
 (n-1) F_n^{\rm sym} = \sum_{m=0}^{n-1} {\rm Sym}[\frac{k}{K_1} G_m F_{n-m}]  =  \frac{1}{2} \sum_{m=0}^{n-1} {\rm Sym}[\frac{k^2}{K_1 K_2} G_m G_{n-m}],
 \end{equation}
noting that $k^2/(K_1K_2) = k/K_1 + k/K_2$, it follows that
\begin{equation}
X_n^{\rm sym} = \frac{1}{2n+3}F_n^{\rm sym}; ~~~ Y^{\rm sym}_n = \frac{2}{2 n+3} F_n^{\rm sym},
\end{equation}
where $X_n$ and $Y_n$ were defined in eqs.~(\ref{eqn:Xn}) and (\ref{eqn:Yn}).
It is trivially seen that these satisfy the symmetrized SPT recursion relations, eq.~(\ref{eqn:SPT_rec}). $\blacksquare$

\subsection{Derivation of SPT from LPT}
\label{app:LPTtoSPT}

To generate SPT-like expressions for $P(k)$ from LPT, $P_{\rm LPT}(k)$ can be expanded in powers of $P_L(k)$:
\begin{eqnarray}
  P_{\rm LPT}(k) &=& \int dq\, e^{i k \,q}
  \left(-\frac{k^2}{2}\sigma^2(q) +
  \frac{k^4}{8}\sigma^4(q)+ \cdots\right), \label{eqn:Pksigexpa} \\
  &=& P_L + \frac{1}{8} \int dq\, e^{i k \,q}\,\nabla_q^4
  \,\sigma^4(q) + {\cal O}(P_L^3), \\
  &=& P_L + \frac{1}{8} \int dq\, e^{i k \,q} 
  \left[\vphantom{\int} 6 \,([\sigma^2]'')^2 + 8\,([\sigma^2]' [\sigma^2]''') +
  2\, [\sigma^2] [\sigma^2]'''' \right] + {\cal O}(P_L^3),\label{eqn:Pksigexp2}\\
 &=&  \left(1 -k^2 \eta^2\right) P_L + \int \frac{dk'}{2\pi}
\left [3  + 4 \frac{k-k'}{k' } + \frac{(k-k')^2}{k'^2} \right ] P_L(k')P_L(k-k') \nonumber \\
&&+   {\cal O}(P_L^3),  \label{eqn:pkLPTPkexpAp}
\end{eqnarray}
where the second line used that $k^4 e^{ik\,q}=\nabla_q^4 e^{ik\,q}$,
which integrating by parts equals $e^{ik\,q}\nabla_q^4$ up to surface terms,
and the last line substitutes for $\sigma^2$ using eq.~(\ref{eqn:sigmaq}).
As anticipated, the last line in the above equation is identical to what we found for 1-loop SPT (eq.~\ref{eqn:SPT1loop}), and Appendix~\ref{ap:proof} proves that this correspondence holds at all loop orders. 

Similarly, we can also start with the correlation function in the LPT (eq.~\ref{eqn:xiZeld}) and expand:
\begin{eqnarray}
1 + \xi_{\rm ZA}(x) &=& \int \frac{dq}{\sqrt{2\pi} \sigma(q)} \exp \left[-\frac{(q-x)}{2\sigma^2(q)} \right], \\ 
            & = & \int \frac{dx'}{\sqrt{2\pi}} \exp \left[-\sum_{m=0}^\infty \frac{x'^m}{m!} \partial_y^m \left\{\frac{x'^2}{2 \sigma(x +y)^2} + \frac{1}{2}\log[\sigma(x +y)^2] \right\} \bigg |_{y=0} \right], \nonumber\\
            &= & \left \langle  \sum_{j=0}^\infty \frac{x'^j}{j!}  \left[-\sum_{m=1}^\infty \frac{x'^m}{m!} \partial_y^m \left\{\frac{x'^2}{2 \sigma(x +y)^2} + \frac{1}{2}\log[\sigma(x +y)^2] \right\} \bigg |_{y=0} \right]^j \right \rangle_{\sigma}, \nonumber 
\end{eqnarray}
where $\langle\cdots\rangle_{\sigma}$ denotes an average over $x'$ with a Gaussian kernel of width $\sigma$.  The last line allows us to evaluate each term in the expansion using Wick's theorem $\langle x'^{2m} \rangle_{\sigma} = (m-1)!! \sigma^{2m}$, which reduces to
\begin{eqnarray}
\xi_{ZA}(x) &=& \frac{[\sigma^2]''}{2} + \frac{3}{4} ([\sigma^2]'')^2 + ([\sigma^2]' [\sigma^2]''') + \frac{1}{4} [\sigma^2] [\sigma^2]'''' + {\cal O} \left( \frac{\sigma}{x} \right)^6,\\
 &=& \xi_{L}(x) + 3 \xi_L(x)^2 + 4 \xi_L(x)' \int^\infty_x dx \xi_L(x) + \frac{\sigma^2}{2}  \xi_L(x)''+  {\cal O} \left( \frac{\sigma}{x} \right)^6. \label{eqn:xiexpanap}
\end{eqnarray}
Primes denote derivatives with respect to the argument, and for the error term we have treated, e.g., $[\sigma^2/x^2]^3 \sim [{\sigma^2}'']^3$ as equivalent.  In addition, we have used that $\xi_{L} =  [\sigma^2]''/{2}$.  Eq.~(\ref{eqn:xiexpanap}) is identical to our 1-loop expression for the correlation function (eq.~\ref{eqn:xiexpan}), and it is the Fourier transform of eq.~(\ref{eqn:pkLPTPkexpAp}).  

\subsection{Stochastic fluctuations in standard theories}
\label{app:stochastic}

At large scales it is well known that the nonlinear evolution of a power spectrum that has no power as $k\rightarrow 0$ is to develop a tail in which $P(k) \sim k^{4}$ plus terms that are higher in power \cite{peebles}.  This limiting behavior can also be derived from our SPT and LPT expressions.  We have shown that the power spectrum can be expanded to yield  (e.g., eq.~\ref{eqn:Pksigexpa})
\begin{eqnarray}
  P_{\rm 1-loop}(k) &=& \int dq\, e^{i k \,q}
  \left(-\frac{k^2}{2}\sigma^2(q) +
  \frac{k^4}{8}(\sigma_l^2 +  \sigma_s^2)^2(q)+ \cdots\right), \label{eqn:Pksigexp3} \\
  & = & P_L(k) + \frac{k^4}{8}\widetilde{(\sigma_l^2 +  \sigma_s^2)} \star\widetilde{(\sigma_l^2 +  \sigma_s^2)}  (k) +\cdots,
 \end{eqnarray}
 where we have written explicitly the terms up to one-loop order.  The purely stochastic term, $\propto k^4 \widetilde{\sigma_s^2} \star \widetilde{\sigma_s^2}$, scales as $k^4$ as $k\rightarrow 0$ as this term is convolving two broadband high-pass filtered fields (e.g., the convolution does not depend on $k$ in the limit $k \ll k'$).  This behavior is not as apparent in $P_{22}$ in the text (i.e. eq.~\ref{eqn:P22} or equivalently eq.~\ref{eqn:pkLPTPkexpAp}).  

\section{Equations in the effective field theory of large-scale structure}

\subsection{Derivation of smoothed equations of motion}
\label{app:stresstensor}

The presentation in this section follows the derivation of effective equations of motion in \cite{baumann12} and \cite{carrasco12}, but also comments on some recent improvements to these original derivations.  We start with the collisionless Boltzmann equation for the particle distribution function $f(p, r)$, where $p$ is the momentum including the Hubble flow contribution and $r$ is the proper coordinate:
\begin{equation}
\frac{df(r, p, t)}{dt} = \frac{\partial f}{\partial t} + \frac{p}{m} \nabla_r f  - m \nabla_r \Phi  \nabla_p f =0,
\end{equation}
where includes the homogeneous component so, e.g., $\Phi= \phi +H^2r^2/4$ in Einstein de-Sitter.
We smooth this equation on a comoving scale $\Lambda^{-1}$, again using the shorthand $X_l  \equiv \int dr' W_\Lambda(r-r') X(r')$ for the smoothed long wavelength field.  Since our smoothing is a linear operation,
\begin{equation}
 \frac{\partial f_l}{\partial t} +  \frac{p}{m} \nabla_r {f_l}  =   m \int dr' W_\Lambda(r-r')  \nabla_{r'}\Phi(r') \nabla_p f(r', p).
 \label{eqn:smboltz}
\end{equation}
We define the density, momentum, and velocity dispersion fields to be
\begin{eqnarray}
\rho &=& m \int dp\, f(r, p, t),\\
\pi &=& \int dp\, p f(r, p, t),\\
\sigma_v^2 &=&  m^{-1}\int dp\, p^2 f(r, p, t).
\end{eqnarray}
To proceed we take moments of equation~(\ref{eqn:smboltz}) by integrating each term over $\int dp\, p^n$.  Since the right hand side of equation~(\ref{eqn:smboltz}) is a total derivative with respect to $p$, the zeroth moment yields the continuity equation:
\begin{equation}
  \partial_t \rho_l  + \nabla_r \pi_l =   m^2\int dr' W(r-r')[\nabla_{r'}\Phi(r')]\int dp \, \nabla_p f(r',p) = 0,
\end{equation}
where in the last step we have used that the integral of a derivative
is zero.  If we define $v_l^b=\pi_l/\rho_l$ (the superscript on $v_l^b$ is to denote that this is not truly a long-wavelength field), we obtain the continuity
equation in its usual form:
\begin{equation}
  \dot \rho_l + \nabla_r (\rho_l v_l^b)= 0.
\label{eqn:conteff2}
\end{equation}

Next consider the $1^{\rm st}$ moment, which should give an Euler-like
or Navier-Stokes-like equation.  Multiplying by $p$ and integrating
Eq.~(\ref{eqn:smboltz}) over $p$ gives
\begin{equation}
  \partial_t \pi_l + \nabla_r \sigma^2_{v,l} = 
  m\int dr'\ W(r-r')[\nabla_{r'}\Phi(r')]\int dp\, p\,\partial_p f(r',p),
\end{equation}
where $\sigma^2_{v,l}$ is the smoothing of $\sigma_v^2$, not the square of
the smoothed sigma.  Integrating by parts the integral over $dp$ yields
\begin{equation}
  \partial_t \pi_l + \nabla_r \sigma^2_{v,l} =
  -\int dr'\ W(r-r')[\nabla_{r'}\Phi(r')]\rho(r').
\end{equation}
Writing $\pi_l=\rho_l v_l^b$ and, then, pulling out the linear terms from the RHS, we have
\begin{eqnarray}
  \rho_l\partial_t v_l^b + v_l^b\partial_t\rho_l &+& \nabla_r \sigma^2_{v,l}
  = -\int dr'\ W(r-r')[\nabla_{r'}\Phi(r')]\rho(r'),\\
    &\approx & -\nabla_{r} \Phi_l \rho_l +\bar{\rho}  \nabla_{r} \phi_l \delta_l  - \bar \rho \int dr'\ W(r-r')[\nabla_{r'}\phi(r')]\delta(r').\label{eqn:contapprox}
\end{eqnarray}
To reach this equation, we assumed that the homogeneous part of $\Phi$, $\Phi_{h}(r)$, can be approximated as being evaluated at $r$ rather than $r'$ in order to pull it out of the integral.  To calculate the error from this approximation, we expand $\Phi_h$ around $r'$ as $\Phi_h(r') = \Phi_h(r) +  \nabla_r \Phi_h(r) (r' -r) + \frac{1}{2}\nabla_r^2 \Phi_h(r) (r' -r)^2 + ....$ since it is smoothly varying (in Einstein de-Sitter and $\Lambda$CDM this expansion is exact at second order).  Furthermore, we  specialized to the Gaussian functional from for $W(r-r') = \Lambda \exp[- \Lambda^2 (r-r')^2/2]/\sqrt{2 \pi}$ and to an Einstein deSitter cosmology such that
\begin{eqnarray}
 \bar \rho  \int dr'\ W(r-r')\left[\nabla_{r'}\Phi_h(r') - \nabla_r \Phi_h(r)\right]\delta(r') &=&  \bar \rho \int dr'\ W(r-r')[\nabla_r^2 \Phi_h(r) (r' -r)]\delta(r'), \nonumber \\
&=&  -4\pi G \bar \rho^2 \int dr'\ W(r-r')[r' -r]\delta(r'),  \nonumber\\
&=& \frac{3}{2} H^2 \bar \rho \int dr'\ \Lambda^{-2} \nabla_r W(r-r')  \delta(r'), \nonumber \\
&=& -\frac{3}{2} H^2  \bar \rho \int dr' a^2 W(r-r')  \frac{ \nabla_r' \delta(r')}{\Lambda^2}, \nonumber \\
&=&  -\frac{3}{2} H^2  \bar \rho a^2 \frac{\nabla_r \delta_l}{\Lambda^2},
\end{eqnarray}
 where $\Lambda$ is comoving.  Thus, the error from approximation is suppressed by $\Lambda^2$, justifying our approximation.

Starting again with eq.~(\ref{eqn:contapprox}), we can use the continuity equation to eliminate $\partial_t \rho_l$, which yields:
\begin{eqnarray}
  \partial_t v_l^b -  \rho_l^{-1} v_l^b\nabla_r(\rho_l v_l^b)
  + \rho_l^{-1}\nabla_r \sigma^2_{v,l} + \nabla_{r} \Phi
  &=& -\rho_l^{-1} \left(\bar \rho \left[\nabla_r \phi \delta\right]_l -\bar{\rho}  \nabla_{r} \phi_l \delta_l  \right).
  \end{eqnarray}
Furthermore, we can use the Poisson equation, $\nabla_r^2 \phi = -4\pi G \bar{\rho} \delta$, to eliminate $\delta$ on the RHS:
\begin{eqnarray}
  \partial_t v_l^b + v_l^b\nabla_r v_l^b  -  \rho_l^{-1} \nabla_r(\rho_l {v_l^b}^2)
  + \rho_l^{-1}\nabla_r \sigma^2_{v,l}  + \nabla_{r} \Phi_l
  &=& \frac{1}{4\pi G \rho_l} \left(\left[\nabla_r \phi \nabla_{r}^2 \phi\right]_l -   \nabla_{r} \phi_l \nabla_{r}^2 \phi_l \right),\nonumber
  \end{eqnarray}
 or
 \begin{eqnarray}
  \partial_t v_l^b + v_l^b\nabla_r v_l^b  + \nabla_{r} \Phi_l
  + \frac{\nabla_r(\sigma_{v,l}^2-\rho_l {v_l^b}^2)}{\rho_l}
  &=& \frac{1}{8\pi G \rho_l}\nabla_r  \left(\left[\nabla_r \phi \nabla_{r} \phi\right]_l -   \nabla_{r} \phi_l \nabla_{r} \phi_l \right),\nonumber
\end{eqnarray}
which can be written as
\begin{equation}
  \partial_t v_l^b + v_l^b \nabla_r v_l^b + \nabla_r \Phi_l
  = -\frac{1}{\rho_l}\nabla_r \tau_\Lambda.
\end{equation}
We decompose the long-wavelength stress tensor as $\tau_\Lambda = \kappa_\Lambda +\Xi_\Lambda$, the sum of a kinematic term,
\begin{equation}
\kappa_\Lambda = \sigma_{v,l}^2 - \rho_l {v_l^b}^2,
\label{eqn:kappal}
\end{equation}
and a gravitational term,
\begin{equation}
\Xi_\Lambda = -\frac{1}{8\pi G} \left([\nabla_r \phi \nabla_r \phi]_l - \nabla_r \phi_l \nabla_r \phi_l\right).
\label{eqn:Phil}
\end{equation}

The above equations are the dynamical equations of EFTLSS.  
 To make better contact with the velocity that appears in standard perturbation theories as well as to be able to measure the coefficients of the effective theory on a grid, we formulate the EFTLSS in terms of the Eulerian velocity $v_l$ (rather than $v_l^{b}$) via an expansion in $\delta_l$:
\begin{eqnarray}
v_l^{b} \equiv \frac{\pi_l}{\rho_l} =  \frac{[(1+\delta) v]_l}{1+\delta_l} &=& v_l + [v (\delta - \delta_l)]_l  - ([v \delta]_l  - v_l \delta_l )  \delta_l +{\cal O}(\delta_l^2),\\
&=&  v_l + [v (\delta - \delta_l)]_l (1-\delta_l)   +{\cal O}(\delta_l^2,  \Lambda^{-2}).
\end{eqnarray}
We note that the Eulerian velocity $v_l$ is still sensitive to short wavelengths as detailed in \cite{carrasco14}, affecting $>1~$loop order predictions (which are not of concern here).

Using this expansion, the continuity equation becomes, to linear order in the long-wavelength fields,
\begin{eqnarray}
  \dot \rho_l + \nabla_r (\rho_l v_l)&=&  -\nabla_r  \left\{\rho_l [v (\delta - \delta_l)]_l -  \rho ([v \delta]_\Lambda  - v_l \delta_l )  \delta_l \right \} + {\cal O}(\delta_l^2),\\
  &=& -\bar \rho \; \nabla_r [v (\delta - \delta_l)]_l + {\cal O}(\delta_l^2,  \Lambda^{-2}).
\label{eqn:conteff2}
\end{eqnarray}
Where the latter relation holds to the extent that large-scale fields do not affect small-scale averaging (i.e. it drops terms suppressed by $\Lambda^{-2}$).  Nicely, we now have terms on the RHS of the continuity equation that in effect generate terms that \cite{mercolli14} showed are not forbidden by symmetry and that are needed to cancel divergences that appear in the 1-loop velocity power spectrum or higher loop calculations.  

Turning to the Euler equation and plugging in our expansion for $v^b_l$:
\begin{equation}
  \partial_t v_l + v_l \nabla_r v_l + \nabla_r \Phi_l
  = -\bar \rho^{-1} \nabla_r \tau_\Lambda -  \partial_t   \{[v (\delta - \delta_l)]_l (1-\delta_l)\}   - \nabla_r \left( v_l  [v (\delta - \delta_l)]_l (1-\delta_l) \right) + {\cal O}(\delta_l^2,  \Lambda^{-2}).
\end{equation}
Note that, e.g., $[\nabla v (\delta - \delta_l)]_\Lambda$ can contribute a constant and so we maintain terms that are higher order.  The above physical coordinate equations can all be written as the comoving coordinate equations that are solved perturbatively in SPT (e.g., \cite{peebles}; $\nabla_r \rightarrow a^{-1} \nabla$ and $\partial_t = \partial_t - H x \nabla$). The only difference with respect to SPT is that the terms with $\tau_\Lambda$ and $[v (\delta - \delta_l)]_l$.  These additional terms are sensitive to short-wavelength perturbations that are nonlinear in the cosmologies we are considering.

What we need to solve for the nonlinear density is just the nonlinear equation for $\delta$, which plugging in our new Euler and continuity equations and going to comoving coordinates yields
\begin{eqnarray}
&&-a^2 {\cal H}^2(a) \partial_a^2 \delta_l -a \left(2 {\cal H}^2(a) + a {\cal H}(a) d{\cal H}(a)/da \right)   \partial_a \delta_l  + 4 \pi G \bar{\rho} a^2 \delta_l  = \left( a {\cal H} \partial_a + {\cal H} \right)  \nabla (\delta_l u_l) -\nabla (u_l \nabla u_l) \nonumber\\
&&\underbrace{  -\bar \rho^{-1} \nabla^2  \tau_\Lambda + \nabla a \calH \partial_a ( [u(\delta - \delta_l)]_l \delta_l) - \nabla^2 \left( [u (\delta - \delta_l)]_l u_l \right)- \nabla \calH [u (\delta - \delta_l)]_l (1-\delta_l)}_{\rm terms~not~in~SPT} +  {\cal O}(\delta_l^2, \Lambda^{-2}). \nonumber
\label{eqn:nonlinear2}
\end{eqnarray}

 To solve the EFTLSS equations perturbatively, the new short-wavelength sensitive terms are expanded in terms of the long-wavelength fields (at all previous times; \cite{carroll14, carrasco14}).  The coefficients of these terms are then estimated from a simulation in Appendix~\ref{sec:stochastic}.  Another approach to deriving the effective equations is to include all possible terms allowed by the symmetries of the problem -- homogeneity \& isotropy, Galilean invariance, mass and momentum conservation, and the equivalence principle -- as ``source'' terms to the continuity and Euler equations \cite{mercolli14}.  A quick summary in the spirit of this approach is described in the ensuing appendix.

\subsection{Symmetries approach}
\label{ap:EFTsymmetry}

In section \ref{sec:EFTLSS} we described one method for overcoming the
deficiencies of SPT.  In that approach, an attempt was made to construct
dynamical equations for the smoothed matter overdensity and velocity fields.
An alternative approach \cite{mercolli14}, which arrives at the same extra terms as we find in eq.~(\ref{eqn:1loop}) is to write down all of
the terms consistent with the symmetries.  (We note that the symmetries are also taken account in the first method; this method is just more pedagogical.)  We briefly review this approach here.

When computing ensemble averages of cosmological fields we can imagine
taking the average in two steps.
First we average over the small-scale (and potentially nonlinear)
fluctuations and then over the longer wavelength modes (for which we
assume perturbation theory is applicable).
In the intermediate stage, after ``integrating out'' the small-scale
perturbations, we have a set of evolution equations for the long-wavelength
modes in which the effects of the small-scale perturbations are encoded
as additional terms beyond the terms one usually encounters in perturbation
theory.
If we write the effective equations for the long-wavelength perturbations
with the (standard) linear terms of the left-hand side
(e.g.~eqs.~\ref{eqn:effcont} and \ref{eqn:effnavierstokes}),
then the right hand side must contain all other terms allowed by the
symmetries of the problem, though with unknown coefficients.
In addition to the standard quadratic terms [$\nabla(\delta u)$ and
$\nabla(u\nabla u)$] we must handle extra terms such as those in
$\tau_\Lambda$ of eq.~(\ref{eqn:effnavierstokes}).  One can write all such terms consistent with the symmetries to a given order in $\delta$, $u$, and derivatives, with the full effective equations having infinitely many terms.

To make progress these equations can be solved perturbatively, expanding in the linear overdensity, $\delta_L$, and only treating terms in the equations at or below the desired order in $\delta$, $u$, and derivatives.
We arrange the source terms in a derivative expansion
(i.e.~in Fourier space an expansion in positive powers of $k^2$)
and also a Taylor expansion in powers of the long-wavelength perturbations
(upon which the small-scale dynamics can depend).
At lowest order, the only dependence
can be on $\delta_L$ since at linear order $\theta_L=\delta_L$ and the
vorticity vanishes.  At the next nontrivial order and in Fourier space the allowed terms go
as a constant times $k^2\delta_L$ -- it cannot be a constant times $\delta_L$ or mass would not be conserved nor times $k \delta_L$ or momentum conservation would be violated -- or are independent of $\delta_L$ with RMS scaling
as $k^2$ due to mass and momentum conservation \cite{peebles}. These terms will have unknown coefficients with unknown time-dependence.   In the main body, we justify why these terms should be thought of as third order terms. 
When integrated against the Green's function these new terms generate contributions
to the density going as $\delta_l$-independent ($J$) or as a coefficient
times $k^2$ times the linear theory overdensity.

When computing the power spectrum of $\delta_l$ the first term only comes
in as $P_J$ in eq.~(\ref{eqn:1loop}).  We know that $P_J$ goes as $k^4$ as
$k\to 0$ but the behavior outside this limit is unknown.
  The other term contributes $k^2P_L(k)$ to lowest order, with an unknown coefficient
($2\alpha_{c,\Lambda}$ in eq.~\ref{eqn:1loop}) that is typically fit for.  The final result for the density power spectrum
is the `usual' perturbative solution plus a stochastic term
(going as $k^4$ as $k\to 0$) plus a term going as $k^2P_L(k)$.
This is identical to eq.~(\ref{eqn:1loop}).

\section{Importance of stochastic terms}
\label{sec:stochastic}

\begin{figure}
\begin{center}
\epsfig{file=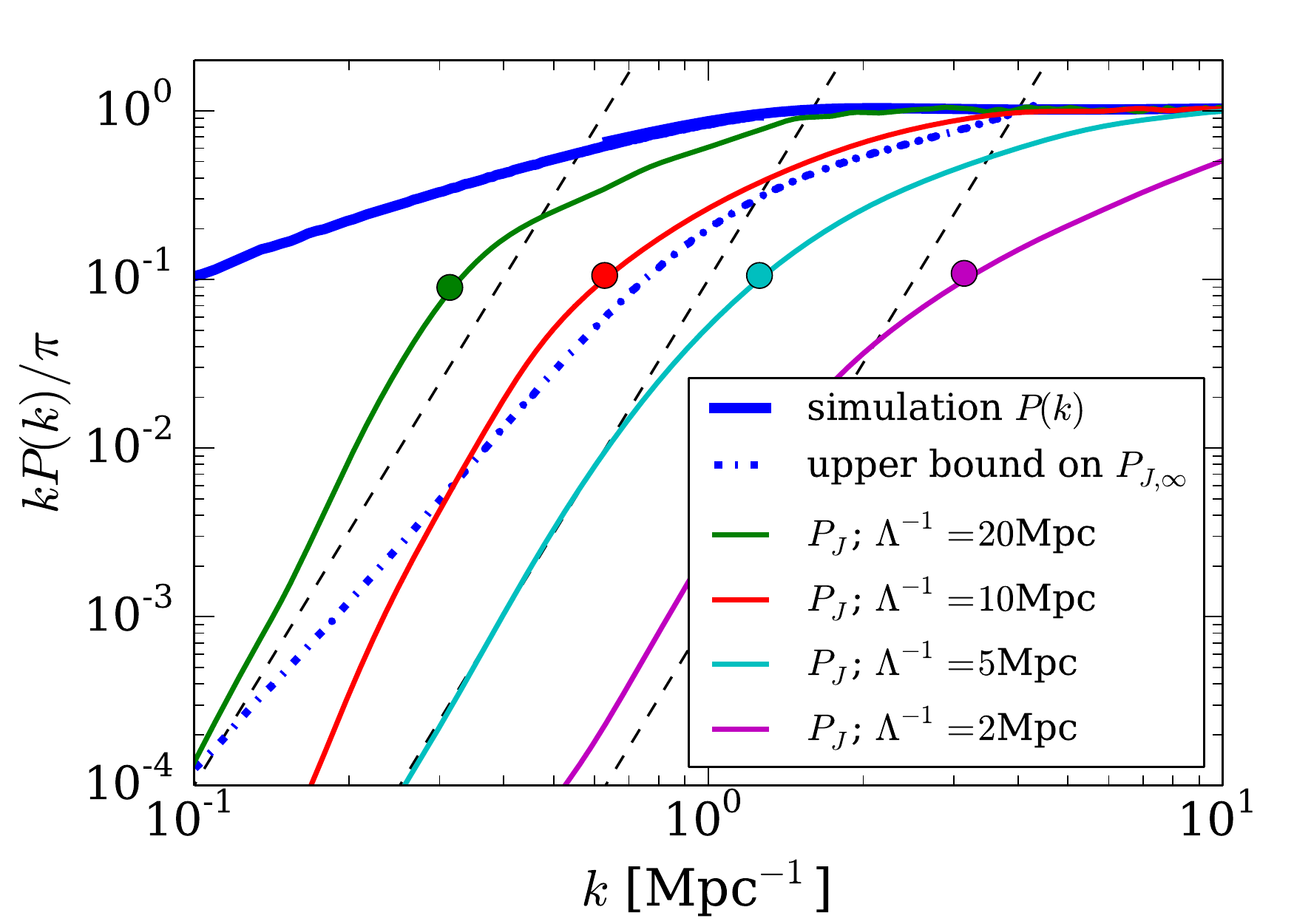, width=10cm}
\end{center}
\caption{Contribution of stochastic (small-scale) fluctuations to the power spectrum at $z=1$ for our CDM-like case.  The stochastic term is computed by differencing $\delta$ in two simulations that have the same modes with $k < 2\pi \Lambda$ and different modes at larger $k$, and then dividing by two.  Interestingly, the curves quickly reach the $k^5$ small wavenumber limit once $k< 2\pi \Lambda$.  The dashed lines show the $k^5$ asymptote that these curves should reach as $k\rightarrow 0$.  The stochastic term is controversially dropped in previous EFTLSS calculations, but the quick asymptote of this term to $\sim k^5$ suggests that dropping it is indeed an excellent approximation.  The dot dashed curve shows a possible upper bound on $P_J$ discussed in the text. \label{fig:stochastic}}
\end{figure}

Previous calculations of $P(k)$ in EFTLSS have dropped the stochastic term $P_J$ in equation~(\ref{eqn:1loop}) -- the term that does not correlate with any term in the expansion in the smoothed density and velocity -- because of its weaker $k\to 0$ scaling \cite{carrasco12, carrasco14}.
This is curious, as halo models for large-scale structure explain nonlinear evolution with just a stochastic term (scaling with the abundance of halos) that goes as $k^0$ \cite{seljak00, ma00, cooray02}.
That halo models are at least qualitatively successful suggests that the
stochastic term may have a non-trivial dependence on wavenumber at mildly
nonlinear wavenumbers, which may invalidate our argument for its smallness
based purely on the $k\rightarrow 0$ limit.
This objection was raised in \cite{mohammed14, seljak15}, which used the halo profile to motivate a completely different expansion from that in EFTLSS for additional contributions to $P(k)$ (although we note that their corrections may not be ``stochastic'' in our sense, i.e.~meaning that they do not correlate with the smoothed fields).  Naively, the impact of the stochastic term and the profiles of nonlinear systems on $P(k)$ should be at least as important in 1D as in 3D because (1) even though their profile may be different, virialized systems still should have the same extent (half the turnaround radius),\footnote{Virialized systems are much less dense in 1D than halos in 3D.  However, we do not expect that the density in the virialized region matters for the power spectrum at mildly nonlinear scales.  Imagine taking a kernel and convolving the density field on a scale of a few Mpc (small enough to not affect mildly nonlinear scales).  This dramatically reduces the density contrast in the field.   However, the power spectrum on mildly nonlinear scales is not affected since it is multiplied by the Fourier transform of the smoothing kernel squared.  Thus, for the matter evolution on mildly nonlinear scales it does not matter what the structure of systems is on less than a few Mpc scales, larger than the sizes of all halos in 3D concordance cosmology.} and (2) there are fewer realizations of the small-scale modes in 1D than in 3D.  

However, we find dropping the stochastic term to be an excellent approximation (and accurate at the percent level a factor of $\sim 3$ beyond the smoothing scale).
Figure~\ref{fig:stochastic} shows the contribution of stochastic (small-scale) fluctuations that are ignored in EFTLSS to the $z=1$ matter power spectrum for our CDM-like case.  The stochastic power, $P_{J}$, is computed by differencing the $\delta$ in simulations that have the same random numbers for modes with $k < 2\pi \Lambda$ and different ones for $k > 2\pi \Lambda$, and then dividing by two.  We note that $\Lambda$ in this case is not the smoothing scale but rather the coherence scale of the two simulations.  The quantity that enters into the EFTLSS power spectrum for finite $\Lambda$, $P_{J,\Lambda}$ is essentially $W_\Lambda^2$ times what is plotted at $k<2 \pi \Lambda$.  Beyond $2\pi \Lambda$ (the filled circles), there is an intermediate range of scales where the stochastic terms scale as $k$ in $k P(k)$ that is most apparent for the smaller $\Lambda$ values in figure~\ref{fig:stochastic} (as in the halo model).  However, the curves quickly reach the $k^5$ large-scale limit, and hence the halo model ansatz that the shape of halos are an important contribution to the mildly nonlinear power is not correct, at least in 1D.  

The $P_J$ curves in figure~\ref{fig:stochastic} do not give an estimate for the actual value for $P_{J,\Lambda}$ as $\Lambda \rightarrow \infty$.  Instead, it is how quickly these curves reach the $k^5$ asymptote that is of most interest, as the steep asymptotic scaling is what is used to drop these terms in current applications of EFTLSS.  In EFTLSS these stochastic contributions should sum, at lowest order, with UV sensitive terms in $P_{22,\Lambda}$ to generate a $\Lambda$-independent term, but their functional form potentially matters near $\Lambda$.  A way to place an upper bound on $P_J$ as $\Lambda \rightarrow \infty$ (to the extent that $P_J$ in this limit makes sense at all) is to take the square of the difference between the nonlinear density field and the Zeldovich approximation density field, normalizing by a transfer function so that Zeldovich has been rescaled to have the same power spectrum:  i.e., to calculate  $\langle (\delta_{NL} - T \delta_{ZA} )^2 \rangle/2$, where $T = ({P_{NL}}/P_{ZA})^{1/2}$ and $\delta_{NL}$ and $P_{NL}$ are the nonlinear density and power spectrum computed in the simulation.  This quantity, which also equals $P_{NL} (1- r^2)$ where $r$ is the cross correlation coefficient between $\delta_{NL}$ and $\delta_{ZA}$, is the part of the nonlinear density that does not correlate with the Zeldovich density and hence more or less excludes the SPT terms at all orders.  It is comprised of both the stochastic term but also non-stochastic contributions from effective terms that do not trace $\delta_{ZA}$.
This quantity is shown by the dot-dashed curve, which is down by three orders of magnitude from the simulation power at $k=0.1~$Mpc$^{-1}$.  The same statistic is similarly suppressed in 3D \cite{tassev13}.

\subsection{Estimating the new coefficients of the effective theory}
\label{sec:estcs}

\begin{figure}
\begin{center}
\epsfig{file=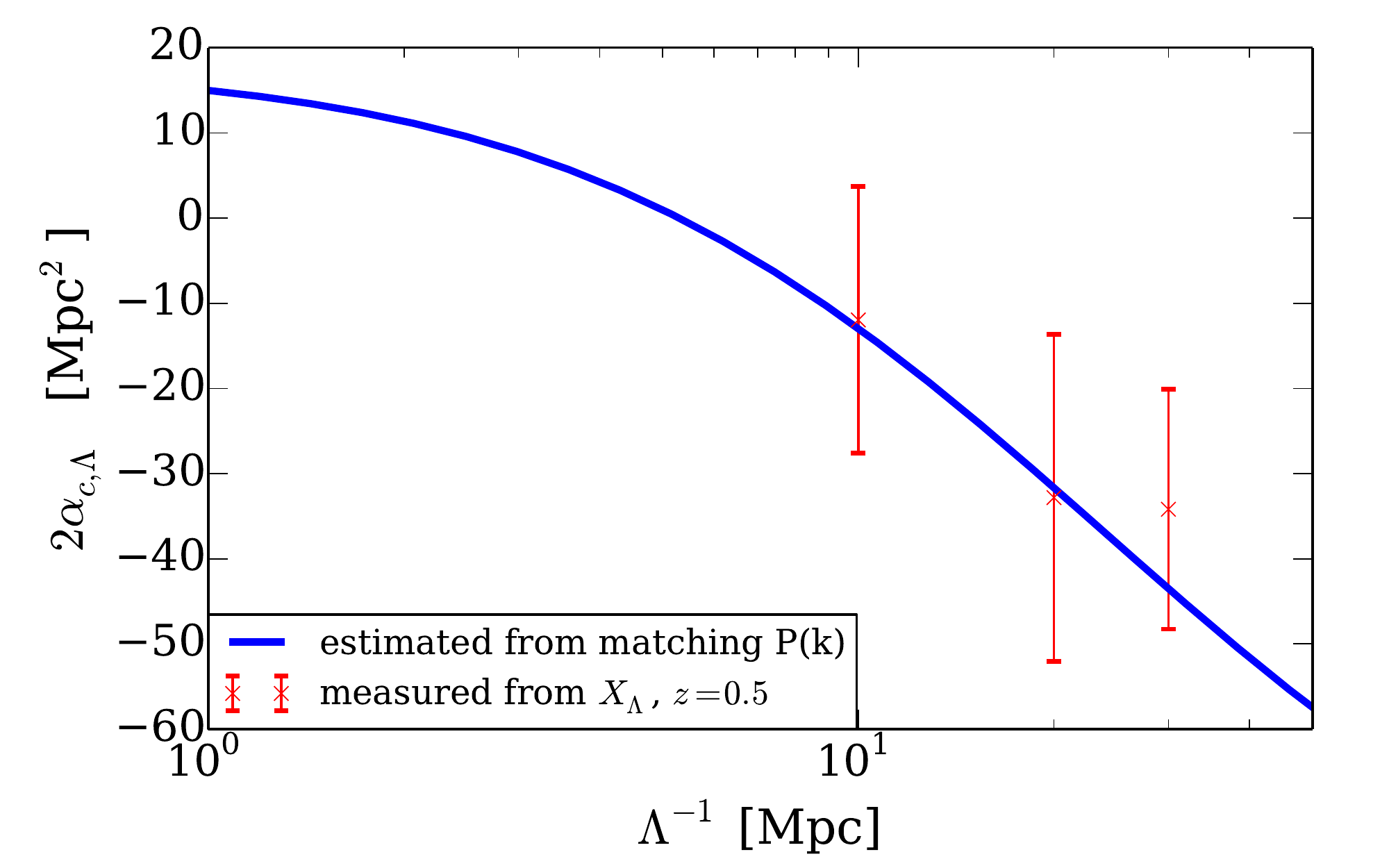, width=10cm}
\end{center}
\caption{Estimates for $2 \alpha_{c,\Lambda}$ constructed by first estimating $c_{\rm tot}^2$ from nonlinear simulations (using eq.~\ref{eqn:ctotest}) and, then, assuming that $c_{\rm tot}^2 \propto a$ to solve for $2 \alpha_{c,\Lambda}$ via eq.~(\ref{eqn:alphaED}).    The solid blue curve corresponds to the $2 \alpha_{c,\Lambda}$ needed to fit the power spectrum in our simulations, using the $2 \alpha_{c}$ estimated in section~\ref{ss:testEFTLSS}.   The points with error bars show the estimated values from $X_\Lambda$.  All curves and points are computed for $z=0.5$.
 \label{fig:alpha}}
\end{figure}

Equation~(\ref{eqn:tau}) for $X_\Lambda$ provides the dependence of the smoothed fields in terms of a ``stress tensor'' that is sensitive to small scales, which we then expanded in terms of large-scale modes and effective coefficients.  By measuring $X_\Lambda$ and how it correlates with large-scale modes, we can measure the coefficients of this theory and thus not have to fit free parameters when we compare the EFTLSS predictions to the power spectrum in simulations.  An estimator for the relevant linear combination of these coefficients, $c_{\rm tot}^2$, from the small-scale behavior of a simulation is 
\begin{equation}
\bar{\rho} ~ \widehat{c_{\rm tot}^2}  = \left(\sum_k w_k  |\tilde \delta_{L,l}(k)|^2 \right)^{-1} \sum_k w_k \tilde X_\Lambda(k)  \tilde \delta_{L,l}^*(k),
\label{eqn:ctotest}
\end{equation}
where the sum should be over  $k \ll \Lambda$ and we choose the weightings, $w_k$, that maximize the $S/N$ assuming Gaussianity.  We compute $X_\Lambda(k)$ from the Eulerian velocity and overdensity field on our 1D simulation grid.  The new EFTLSS parameter that enters into the 1-loop power spectrum, $\alpha_{c, \Lambda}$, is  then related to $c_{\rm tot, \Lambda}$ via eq.~(\ref{eqn:alphac}). 
 
  This procedure has been found to produce the $2\alpha_c$ needed in \cite{carrasco12} using a somewhat different method to the above estimator, although they did choose what seemed like a very fine-tuned range of separations of $7-10~$Mpc for $\Lambda^{-1}=3~$Mpc ($13-18~$Mpc for $\Lambda^{-1}=6~$Mpc).  A variant of this procedure where the time evolution of $\tau_\Lambda$ is measured explicitly (without assuming a long-wavelength expansion) was also found to work in \cite{manzotti14}.

Figure~\ref{fig:alpha} shows our attempts to measure $c_{\rm tot}^2$ from a 1D simulation in our CDM-like cosmology at $z=0.5$ for different $\Lambda$, where what is plotted is $2 \alpha_{c,\Lambda}$, which is proportional to $c_{\rm tot}^2$. The proportionality constant is determined using eq.~(\ref{eqn:ctotest}) and $c_{\rm tot}^2 \propto a$ as required to scale with $P_{13}$, but it is likely the true temporal scaling is somewhat different and so these estimates are only approximate.  The solid curve corresponds to the $2 \alpha_{c,\Lambda}$ required to match the power spectrum in simulations, determined in section~\ref{ss:testEFTLSS}.   The points with error bars show the estimated value from $X_\Lambda$ (red crosses).  
  These points are in agreement with the simulation fit.  The error bars are rather large, which owe to difficulties in estimating the Eulerian space velocity that appears.   (Indeed, these calculations required a higher density of sheets per cell than our fiducial simulations for the error bars do be constraining at all.  This computation was done with $10^9$ sheets and $10^7$PM grid cells in a $10^6\,$Mpc box.)   It likely reduces the variance to formulate $X_\Lambda$ in terms of the momentum, as done in \cite{carrasco12}.  

\bibliographystyle{JHEP}
\bibliography{References}
\end{document}